\documentclass[12pt]{article}
\usepackage{enumerate}
\usepackage{amsfonts}
\usepackage{amsmath}
\usepackage{amssymb}
\usepackage{amsthm}
\usepackage{latexsym}
\usepackage{color}

\def\H{\mathcal{H}}

\def\M{\mathbb{M}}
\def\P{\mathcal{P}}
\def\S{\mathfrak{S}}
\def\C{\mathfrak{C}}

\def\F{\mathfrak{F}}
\def\T{\mathfrak{T}}
\def\B{\mathfrak{B}}

\newcommand{\supp}{\mathrm{supp}}
\newcommand{\rank}{\mathrm{rank}}
\newcommand{\id}{\mathrm{Id}}
\newcommand{\Tr}{\mathrm{Tr}}

\newcommand{\djn}{\mathrm{dj}}
\newcommand{\shs}{\hspace{1pt}}
\newcommand{\hsh}{\hspace{-1pt}}
\newcounter{defin}  \newcounter{lemma}  \newcounter{theorem}
\newcounter{property} \newcounter{corol}  \newcounter{remark} \newcounter{example}
\newenvironment{lemma}{\par\refstepcounter{lemma}     \textbf{Lemma \thelemma.} }{\rm\par}
\newenvironment{theorem}{\par\refstepcounter{theorem}     \textbf{Theorem \thetheorem.}\ }{\rm\par}
\newenvironment{property}{\par\refstepcounter{property}     \textbf{Proposition \theproperty.}\ }{\rm\par}
\newenvironment{corollary}{\par\refstepcounter{corol}     \textbf{Corollary \thecorol.} }{\rm\par}
\newenvironment{definition}{\par\refstepcounter{defin}     \textbf{Definition \thedefin.}\ }{\rm\par}
\newenvironment{remark}{\par\refstepcounter{remark}     \textbf{Remark \theremark.}}{\rm\par}
\newenvironment{example}{\par\refstepcounter{example}     \textbf{Example \theexample.}}{\rm\par}
\begin{document}

%\title{Convergence conditions for the quantum relative entropy and other applications of the deneralized Dini-type lemma}

\title{Lower semicontinuity of the relative entropy disturbance and its corollaries}

%\title{Quantum relative entropy: general convergence criterion and preservation of convergence under CP linear maps}

\author{M.E. Shirokov\footnote{email:msh@mi.ras.ru}\\Steklov Mathematical Institute, Moscow}
\date{}
\maketitle

%UDC: 519.248.3

%MSC: 81P45, 94A17, 46L53

\begin{abstract} It is proved that the decrease of the quantum relative entropy
under action of a quantum operation is a lower semicontinuous function of a pair of its arguments.
This property implies, in particular, that the local discontinuity jumps  of the quantum relative entropy
do not increase under action of quantum operations. It implies also the lower semicontinuity of the modulus
of the joint convexity of the quantum relative entropy (as a function of ensembles of quantum states).

Various corollaries  and applications of these results are considered.
\end{abstract}

%\emph{Keywords:} Hilbert space, trace-class operator, quantum state, lower semicontinuous function, quantum operation,  strong convergence of quantum operations

\tableofcontents

\section{Introduction}

The quantum relative entropy is one of the basic characteristics of quantum states, which is used essentially in
the study of  information and statistical properties of quantum systems
and channels  \cite{Ume,IRE-1,IRE-2,O&P,W}.\smallskip

From the mathematical point of view,  the quantum relative entropy $D(\rho\|\shs\sigma)$ is a  jointly convex function of a pair $(\rho,\sigma)$ of quantum
states (or, more generally, positive trace class operators) taking values in $[0,+\infty]$. One of the fundamental properties of the quantum relative entropy is its monotonicity
under action of quantum  operations (completely positive
trace-non-increasing linear maps), which means that
\begin{equation*}%\label{m-prop+}
D(\Phi(\rho)\|\shs \Phi(\sigma))\leq D(\rho\shs\|\shs\sigma)
\end{equation*}
for an arbitrary  quantum operation $\Phi:A\to B$ and any quantum states $\rho$ and $\sigma$ on $\H_A$  (with possible value $+\infty$ in one or both sides) \cite{L-REM}.
%Recently, it was shown that this property is valid for any trace-non-increasing positive linear maps $\Phi$ \cite{Reed}.
\smallskip

Another important and widely used property of the quantum relative entropy is its  (joint) lower semicontinuity which means that
the set of all pairs $(\rho,\sigma)$ such that
$D(\rho\shs\|\shs\sigma)\leq c$
is a closed subset of $\T_+(\H)\times\T_+(\H)$ for any $c\geq0$, where $\T_+(\H)$ denotes the cone of positive
trace class operators on $\H$ \cite{O&P,W,L-2}.

In this article we analyse the nonnegative function
$$
\Delta_{\Phi}(\rho,\sigma)\doteq D(\rho\shs\|\shs\sigma)-D(\Phi(\rho)\|\shs\Phi(\sigma))
$$
for a given arbitrary quantum operation $\Phi:A\to B$, which is well defined on the set of all pairs $(\rho,\sigma)$ in $\T_+(\H_A)\times\T_+(\H_A)$ such that $D(\Phi(\rho)\|\shs\Phi(\sigma))$ is finite. The function $\Delta_{\Phi}(\rho,\sigma)$ characterises the degree of reversibility of  a quantum
operation $\Phi$. If $\Delta_{\Phi}(\rho,\sigma)=0$ for some quantum states $\rho$ and $\sigma$ then the Petz theorem (cf.\cite{Petz}) implies that the operation $\Phi$ is (completely) reversible w.r.t. these states, which means the existence of a quantum operation $\Psi=\Psi(\Phi,\rho,\sigma)$ such that $\rho=\Psi\circ\Phi(\rho)$ and $\sigma=\Psi\circ\Phi(\sigma)$.
If $\Delta_{\Phi}(\rho,\sigma)\leq\varepsilon$ then the operation $\Phi$ is $\varepsilon$-reversible w.r.t. the states $\rho$ and $\sigma$ in the sense that
there is a quantum operation $\Psi=\Psi(\Phi,\rho,\sigma)$ such that
$$
F(\rho,\Psi\circ\Phi(\rho))\geq \exp(-\varepsilon/2)\quad \textrm{and} \quad \sigma=\Psi\circ\Phi(\sigma),
$$
where $F(\cdot,\cdot)$ is the fidelity of quantum states \cite{EpsR-1,NQD,EpsR-3}.\smallskip

We will prove that the function $\Delta_{\Phi}(\rho,\sigma)$ is lower semicontinuous on its domain for any given quantum operation $\Phi$. Moreover,
we will prove that the function
$$
(\rho,\sigma,\Phi)\mapsto\Delta_{\Phi}(\rho,\sigma)
$$
is lower semicontinuous on the set
\begin{equation}\label{theset}
\left\{(\rho,\sigma,\Phi)\in\T_+(\H_A)\times\T_+(\H_A)\times\F_{\leq1}(A,B)\,|\,D(\Phi(\rho)\|\shs\Phi(\sigma))<+\infty\right\},
\end{equation}
where $\F_{\leq1}(A,B)$ is the set of all quantum operations from $A$ to $B$ equipped with the strong convergence topology \cite{CSR}.

The last property  means that the set of all triplets $(\rho,\sigma,\Phi)$ such that $\Delta_{\Phi}(\rho,\sigma)\leq\varepsilon$ is a closed
subset of the set in (\ref{theset}) for any $\varepsilon\geq0$. It also implies that the
equality
\begin{equation}\label{RE-D}
D(\rho\shs\|\shs\sigma)=D(\Phi(\rho)\|\shs\Phi(\sigma))+\Delta_{\Phi}(\rho,\sigma)
\end{equation}
is a decomposition of $D(\rho\shs\|\shs\sigma)$ into a sum of two nonnegative lower semicontinuous
functions on the set in (\ref{theset}), since the lower semicontinuity of the function $(\rho,\sigma,\Phi)\mapsto D(\Phi(\rho)\|\shs\Phi(\sigma))$
follows from the lower semicontinuity  of the quantum relative entropy and the definition of the strong convergence.

It follows immediately from decomposition (\ref{RE-D}) that the local continuity of $D(\rho\shs\|\shs\sigma)$ (as a function of
a pair $(\rho,\sigma)$) implies the local continuity of $D(\Phi(\rho)\|\shs\Phi(\sigma))$ and $\Delta_{\Phi}(\rho,\sigma)$
(considered either as functions of a pair $(\rho,\sigma)$ for a fixed operation $\Phi$ or as functions of a triplet
$(\rho,\sigma,\Phi)$). This property was originally proved in \cite{REC} by direct and rather technical way using the convergence criterion for the quantum relative entropy
proposed therein. Moreover, decomposition (\ref{RE-D}) allows us to show that
\begin{equation}\label{D-j}
\limsup_{n\to+\infty}D(\Phi(\rho_n)\|\shs\Phi(\sigma_n))-D(\Phi(\rho_0)\|\shs\Phi(\sigma_0))\leq\limsup_{n\to+\infty}D(\rho_n\|\shs\sigma_n)-D(\rho_0\|\shs\sigma_0)
\end{equation}
for an arbitrary quantum operation $\Phi:A\to B$ and any  sequences  $\,\{\rho_n\}$ and $\{\sigma_n\}$ in $\T_+(\H_A)$ converging to operators $\rho_0$ and $\sigma_0$
such that $D(\rho_0\|\shs\sigma_0)<+\infty$. Since the quantity in the r.h.s. of (\ref{D-j}) characterizes the
local discontinuity of the quantum relative entropy for given converging sequences $\,\{\rho_n\}$ and $\{\sigma_n\}$  (it is always nonnegative and equal
to zero if and only if $D(\rho_n\|\shs\sigma_n)$ converges to $D(\rho_0\|\shs\sigma_0)$), inequality (\ref{D-j}) can be interpreted as a "contraction" of possible discontinuities of the quantum relative entropy by quantum operations.

The essential part of this article is devoted to various applications of the above general
properties to local continuity analysis of the quantum relative entropy and the related functions (the quantum mutual information,
the mean energy, the conditional relative entropy, etc.).

\section{Preliminaries}

\subsection{Basic notations}

Let $\mathcal{H}$ be a separable Hilbert space,
$\mathfrak{B}(\mathcal{H})$ the algebra of all bounded operators on $\mathcal{H}$ with the operator norm $\|\cdot\|$ and $\mathfrak{T}( \mathcal{H})$ the
Banach space of all trace-class
operators on $\mathcal{H}$  with the trace norm $\|\!\cdot\!\|_1$. Let
$\mathfrak{S}(\mathcal{H})$ be  the set of quantum states (positive operators
in $\mathfrak{T}(\mathcal{H})$ with unit trace) \cite{H-SCI,Wilde,BSimon}.

Denote by $I_{\mathcal{H}}$ the unit operator on a Hilbert space
$\mathcal{H}$ and by $\id_{\mathcal{\H}}$ the identity
transformation of the Banach space $\mathfrak{T}(\mathcal{H})$.\smallskip

The \emph{von Neumann entropy} of a quantum state
$\rho \in \mathfrak{S}(\H)$ is  defined by the formula
$S(\rho)=\Tr\eta(\rho)$, where  $\eta(x)=-x\ln x$ if $x>0$
and $\eta(0)=0$. It is a concave lower semicontinuous function on the set~$\mathfrak{S}(\H)$ taking values in~$[0,+\infty]$ \cite{W,L-2,H-SCI}.
The von Neumann entropy satisfies the inequality
\begin{equation}\label{w-k-ineq}
S(p\rho+(1-p)\sigma)\leq pS(\rho)+(1-p)S(\sigma)+h_2(p)
\end{equation}
valid for any states  $\rho$ and $\sigma$ in $\S(\H)$ and $p\in(0,1)$, where $\,h_2(p)=\eta(p)+\eta(1-p)\,$ is the binary entropy \cite{O&P,Wilde}.\smallskip

We will use the  homogeneous extension of the von Neumann entropy to the positive cone $\T_+(\H)$ defined as
\begin{equation}\label{S-ext}
S(\rho)\doteq(\Tr\rho)S(\rho/\Tr\rho)=\Tr\eta(\rho)-\eta(\Tr\rho)
\end{equation}
for any nonzero operator $\rho$ in $\T_+(\H)$ and equal to $0$ at the zero operator \cite{L-2}.\smallskip

By using concavity of the entropy and inequality (\ref{w-k-ineq}) it is easy to show that
\begin{equation}\label{w-k-ineq+}
S(\rho)+S(\sigma)\leq S(\rho+\sigma)\leq S(\rho)+S(\sigma)+H(\{\Tr\rho,\Tr\sigma\})
\end{equation}
for any $\rho$ and $\sigma$ in $\T_+(\H)$, where $H(\{\Tr\rho,\Tr\sigma\})=\eta(\Tr\rho)+\eta(\Tr\sigma)-\eta(\Tr(\rho+\sigma))$
-- the homogeneous extension of the binary entropy to the positive cone in $\mathbb{R}^2$.

The \emph{quantum relative entropy} for two quantum states $\rho$ and
$\sigma$ in $\S(\H)$ is defined as
\begin{equation}\label{URE-def}
D(\rho\shs\|\shs\sigma)=\sum_i\langle
i|\,\rho\ln\rho-\rho\ln\sigma\,|i\rangle,
\end{equation}
where $\{|i\rangle\}$ is the orthonormal basis of
eigenvectors of the state $\rho$ and it is assumed that
$D(\rho\shs\|\sigma)=+\infty$ if $\,\mathrm{supp}\rho\shs$ is not
contained in $\shs\mathrm{supp}\shs\sigma$ \cite{Ume,H-SCI,Wilde}.\footnote{The support $\mathrm{supp}\rho$ of an operator $\rho$ in $\T_+(\H)$ is the closed subspace spanned by the eigenvectors of $\rho$ corresponding to its positive eigenvalues.}\smallskip

Let $H$ be a positive (semi-definite)  operator on a Hilbert space $\mathcal{H}$ (we will always assume that positive operators are self-adjoint). Denote by $\mathcal{D}(H)$ the domain of $H$. For any positive operator $\rho\in\T(\H)$ we will define the quantity $\Tr H\rho$ by the rule
\begin{equation}\label{H-fun}
\Tr H\rho=
\left\{\begin{array}{l}
        \sup_n \Tr P_n H\rho\;\; \textrm{if}\;\;  \supp\rho\subseteq {\rm cl}(\mathcal{D}(H))\\
        +\infty\;\;\textrm{otherwise,}
        \end{array}\right.
\end{equation}
where $P_n$ is the spectral projector of $H$ corresponding to the interval $[0,n]$, ${\rm cl}(\mathcal{D}(H))$ is the closure of $\mathcal{D}(H)$. \smallskip

We will use the following notion introduced in \cite{REC}. \smallskip

\begin{definition}\label{scs-def} A double sequence $\{P^n_m\}_{n\geq0,m\geq m_0}$ ($m_0\in\mathbb{N}$) of finite rank  projectors is \emph{completely consistent}
with a sequence $\{\sigma_n\}\subset\T_+(\H)$ converging to an  operator $\sigma_0$ if
\begin{equation}\label{P-prop}
 P^n_{m}\leq P^n_{m+1}, \qquad \bigvee_{m\geq m_0}P^n_m\geq Q_n,
\end{equation}
where $Q_n$ is the projector onto the support of $\sigma_n$, and
\begin{equation}\label{P-prop+}
P^n_m\sigma_n=\sigma_nP^n_m,\qquad  \rank P^n_m\sigma_n=\rank P^n_m,\qquad \|\cdot\|\,\textrm{-}\!\!\lim_{n\to+\infty}P^n_m=P^0_m
\end{equation}
for all $m\geq m_0$ and $n\geq0$, where the limit in the operator norm topology.\smallskip
\end{definition}

It is essential that  \emph{for any sequence $\{\sigma_n\}\subset\T_+(\H)$ converging to an  operator $\sigma_0$
there exists a double sequence $\{P^n_m\}_{n\geq0,m\geq m_0}$  of finite rank  projectors  completely consistent
with the sequence $\{\sigma_n\}$} \cite[Lemma 4]{REC}.
\smallskip

A finite or countable set  $\{\rho_i\}$ of quantum states
with a  probability distribution $\{p_i\}$ is called (discrete) \emph{ensemble} and denoted by $\{p_i,\rho_i\}$. The state $\bar{\rho}=\sum_{i} p_i\rho_i$ is called  the \emph{average state} of  $\{p_i,\rho_i\}$.  The Holevo quantity of an ensemble
$\{p_i,\rho_i\}$ is defined as
\begin{equation}\label{chi-def}
\chi(\{p_i,\rho_i\})= \sum_{i} p_i D(\rho_i\|\bar{\rho})=S(\bar{\rho})-\sum_{i} p_iS(\rho_i),
\end{equation}
where the second formula is valid provided that $S(\bar{\rho})$ is finite. This quantity gives a upper bound on the amount of classical information that can be obtained from quantum measurements over the ensemble \cite{H-SCI,Wilde}.

For a lower semicontinuous function $f$ on a metric space $X$ and a given
sequence $\{x_n\}\subset X$ converging to a point $x_0\in X$ such that $f(x_0)<+\infty$ we will use the quantity
\begin{equation}\label{dj}
\djn(\{f(x_n)\})\doteq\limsup_{n\to+\infty}f(x_n)-f(x_0)
\end{equation}
characterizing the discontinuity jump of the function $f$ corresponding to the sequence $\{x_n\}$.
The lower semicontinuity of $f$ implies that $\,\djn(\{f(x_n)\})\geq0\,$ and that
$$
\djn(\{f(x_n)\})=0\quad\Leftrightarrow\quad \exists\lim_{n\to+\infty}f(x_n)=f(x_0).
$$

We will use the following simple\smallskip

\begin{lemma}\label{vsl}
\emph{If $f$ and $g$ are lower semicontinuous functions on a metric space $X$ taking values in $(-\infty,+\infty]$
then
$$
\djn(\{f(x_n)\})\leq \djn(\{(f+g)(x_n)\})
$$
for any sequence $\{x_n\}\subset X$ converging to $x_0\in X$ such that $(f+g)(x_0)<+\infty$.} \emph{In particular, if
$$
\lim_{n\to+\infty}(f+g)(x_n)=(f+g)(x_0)<+\infty
$$
then
$$
\lim_{n\to+\infty}f(x_n)=f(x_0)<+\infty.
$$}
\end{lemma}

A \emph{quantum operation} $\Phi$ from a system $A$ to a system
$B$ is a completely positive trace-non-increasing linear map from
$\mathfrak{T}(\mathcal{H}_A)$ into $\mathfrak{T}(\mathcal{H}_B)$.
A trace preserving quantum operation is called  \emph{quantum channel} \cite{H-SCI,Wilde}.   \smallskip
For any  quantum operation  $\,\Phi:A\rightarrow B\,$ the Stinespring theorem implies the existence of a Hilbert space
$\mathcal{H}_E$ and  a contraction
$V_{\Phi}:\mathcal{H}_A\rightarrow\mathcal{H}_B\otimes\mathcal{H}_E$ such
that
\begin{equation}\label{St-rep}
\Phi(\rho)=\mathrm{Tr}_{E}V_{\Phi}\rho V_{\Phi}^{*},\quad
\rho\in\mathfrak{T}(\mathcal{H}_A).
\end{equation}
If $\Phi$ is a channel then $V_{\Phi}$ is an isometry \cite{H-SCI,Wilde}.\smallskip

The mutual information $I(\Phi,\rho)$ of a  quantum channel $\Phi:A\to B$ at a state $\rho$ in $\S(\H_A)$
can be defined as
\begin{equation*}%\label{MI-Ch-def}
I(\Phi,\rho)=I(B\!:\!R)_{\Phi\otimes\id_R(\hat{\rho})}\doteq D(\Phi\otimes\id_R(\hat{\rho})\|\Phi(\rho)\otimes\hat{\rho}_R),
\end{equation*}
where $\bar{\rho}$ is pure state in $\S(\H_A\otimes\H_R)$ such that $\Tr_R\bar{\rho}=\rho$ \cite{H-SCI,Wilde}.\smallskip

Denote by $\F_{\leq 1}(A,B)$ the set of all quantum operations from $A$ to $B$ equipped with the \emph{topology of strong convergence}  defined by the family of seminorms $\Phi\mapsto\|\Phi(\rho)\|_1$, $\rho\in\S(\H_A)$ \cite{CSR}. The strong convergence of a sequence $\{\Phi_n\}$ of operations in $\F_{\leq 1}(A,B)$ to an operation $\Phi_0\in\F_{\leq 1}(A,B)$  means that
\begin{equation*}%\label{star+}
\lim_{n\rightarrow\infty}\Phi_n(\rho)=\Phi_0(\rho)\,\textup{ for all }\rho\in\S(\H_A).
\end{equation*}

We will use the following important result. \smallskip

\begin{lemma}\label{D-A} \cite{D-A} \emph{If a sequence $\{\rho_n\}$ of states converges to a state $\rho_0$ w.r.t. the weak operator topology then
the sequence $\{\rho_n\}$  converges to the state $\rho_0$ w.r.t. the trace norm.}
\end{lemma}\smallskip

\subsection{Lindblad's extension of the quantum relative entropy}

The Lindblad's extension of the quantum relative entropy between positive operators $\rho$ and
$\sigma$ in $\mathfrak{T}(\mathcal{H})$ is defined as
\begin{equation*}%\label{qre-def+}
D(\rho\shs\|\shs\sigma)=\sum_i\langle\varphi_i|\,\rho\ln\rho-\rho\ln\sigma+\sigma-\rho\,|\varphi_i\rangle,
\end{equation*}
where $\{\varphi_i\}$ is the orthonormal basis of
eigenvectors of the operator  $\rho$ and it is assumed that $\,D(0\|\shs\sigma)=\Tr\sigma\,$ and
$\,D(\rho\shs\|\sigma)=+\infty\,$ if $\,\mathrm{supp}\rho\shs$ is not
contained in $\shs\mathrm{supp}\shs\sigma$ (in particular, if $\rho\neq0$ and $\sigma=0$)
\cite{L-2}.\smallskip

If the extended von Neumann entropy $S(\rho)$ of $\rho$ (defined in (\ref{S-ext})) is finite
then
\begin{equation}\label{re-exp}
D(\rho\shs\|\shs\sigma)=\Tr\rho(-\ln\sigma)-S(\rho)-\eta(\Tr\rho)+\Tr\sigma-\Tr\rho,
\end{equation}
where $\Tr\rho(-\ln\sigma)$ is defined according to the rule (\ref{H-fun}) and $\eta(x)=-x\ln x$. \smallskip

The function $(\rho,\sigma)\mapsto D(\rho\shs\|\shs\sigma)$ is nonnegative lower semicontinuous and jointly convex on
$\T_+(\H)\times\T_+(\H)$. We will use the following properties of this function:
\begin{itemize}
  \item for any $\rho,\sigma\in\T_+(\H)$ and $c>0$ the following equalities  hold (with possible values $+\infty$ in both sides):
  \begin{equation}\label{D-mul}
  D(c\rho\shs\|\shs c\sigma)=cD(\rho\shs\|\shs \sigma),\qquad\qquad\qquad\qquad\quad\;
  \end{equation}
  \begin{equation}\label{D-c-id}
  D(\rho\shs\|\shs c\sigma)=D(\rho\shs\|\shs\sigma)-\Tr\rho\ln c+(c-1)\Tr\sigma;
  \end{equation}
  \item for any $\rho,\sigma$ and $\omega$ in $\T_+(\H)$ the following inequalities hold (with possible values $+\infty$ in one or both sides)
  \begin{equation}\label{re-ineq}
  D(\rho\shs\|\shs\sigma+\omega)\leq D(\rho\shs\|\shs\sigma)+\Tr\omega, \qquad\qquad\qquad\qquad\qquad\qquad
  \end{equation}
  \begin{equation}\label{re-2-ineq-conc}
  D(\rho+\sigma\|\shs\omega)\geq D(\rho\shs\|\shs\omega)+D(\sigma\shs\|\shs\omega)-\Tr\omega;\qquad\qquad\qquad\quad\;
  \end{equation}
%  \begin{equation}\label{re-2-ineq-conv}
%  D(\rho+\sigma\|\shs\omega)\leq D(\rho\shs\|\shs\omega)+D(\sigma\shs\|\shs\omega)+ H(\{\Tr\rho,\Tr\sigma\})-\Tr\omega,
%  \end{equation}
%  where $H(\{\Tr\rho,\Tr\sigma\})$ is the extended binary  entropy of $\{\Tr\rho,\Tr\sigma\}$ defined in (\ref{w-k-ineq+});
  \item for any $\rho,\sigma,\omega$ and $\vartheta$ in $\T_+(\H)$ the following inequality holds (with possible values $+\infty$ in one or both sides)
  \begin{equation}\label{D-sum-g}
    D(\rho+\sigma\shs\|\shs \omega+\vartheta)\leq D(\rho\shs\|\shs \omega)+D(\sigma\shs\|\shs \vartheta),
  \end{equation}
  if $\rho\sigma=\rho\vartheta=\sigma\omega=\omega\vartheta=0$ then
  \begin{equation}\label{D-sum}
    D(\rho+\sigma\shs\|\shs \omega+\vartheta)=D(\rho\shs\|\shs \omega)+D(\sigma\shs\|\shs \vartheta).
  \end{equation}
 \end{itemize}

Inequalities (\ref{re-ineq}) and (\ref{re-2-ineq-conc}) are easily proved by using representation (\ref{re-exp}) if the extended von Neumann entropy of the operators $\rho$, $\sigma$  and $\omega$ is finite. Indeed, inequality (\ref{re-ineq}) follows from the operator monotonicity of the logarithm,  inequality (\ref{re-2-ineq-conc}) follows from  inequality (\ref{w-k-ineq+}). In the general case these inequalities can be proved by approximation using Lemma 4 in \cite{L-2}.

Inequality (\ref{D-sum-g}) is a direct corollary of the joint convexity of the relative entropy and identity (\ref{D-mul}). Equality (\ref{D-sum}) follows from the definition \cite{L-2}.

We will use Donald's identity
\begin{equation}\label{Donald}
pD(\rho\|\omega)+\bar{p}D(\sigma\|\omega)=pD(\rho\|p\rho+\bar{p}\sigma)+\bar{p}D(\sigma\|p\rho+\bar{p}\sigma)+D(p\rho+\bar{p}\sigma\|\omega)
\end{equation}
where $\bar{p}=1-p$, valid for arbitrary operators $\rho$, $\sigma$ and $\omega$ in $\T_+(\H)$ and any $p\in[0,1]$ \cite{Donald}.\footnote{In Lemma 2 in \cite{Donald}
it was assumed that $\rho$, $\sigma$ and $\omega$ are (normal) states. The generalization to arbitrary operators in $\T_+(\H)$
can be done by using identities (\ref{D-mul}) and (\ref{D-c-id}).}

\section{The main results}

\subsection{The relative entropy disturbance and its lower semicontinuity}

A basic property of the quantum relative entropy is its monotonicity under quantum operations (completely positive
trace-non-increasing linear maps), which means that
\begin{equation}\label{m-prop}
D(\Phi(\rho)\|\shs \Phi(\sigma))\leq D(\rho\shs\|\shs\sigma)
\end{equation}
for an arbitrary  quantum operation $\Phi:\T(\H_A)\to\T(\H_B)$ and any operators $\rho$ and $\sigma$ in $\T_+(\H_A)$ \cite{L-REM}.\footnote{Here and in what follows $D(\cdot\shs\|\shs\cdot)$ is Lindblad's extension of the Umegaki relative entropy (\ref{URE-def}) to operators in $\T_+(\H_A)$ described in Section 2.2.}\smallskip

Monotonicity property (\ref{m-prop}) means the nonnegativity of the function
$$
\Delta_{\Phi}(\rho,\sigma)\doteq D(\rho\shs\|\shs\sigma)-D(\Phi(\rho)\|\shs\Phi(\sigma))
$$
well defined on the set of all pairs $(\rho,\sigma)$ in $\T_+(\H_A)\times\T_+(\H_A)$ such that $D(\Phi(\rho)\|\shs\Phi(\sigma))$ is finite.
This function can be called the \emph{relative entropy disturbance} by $\Phi$.\footnote{It is also called the relative entropy difference \cite{NQD,EpsR-3}.}

It turns out that the function $\Delta_{\Phi}(\rho,\sigma)$ appears (explicitly or implicitly) in different tasks of quantum information theory. To show this it suffices to
consider the following "special realizations" of this function:
 \begin{itemize}
   \item if $\rho$ is a state in $\S(\H_{ABC})$, $\sigma=\rho_A\otimes\rho_{BC}$ and $\Phi=\Tr_B(\cdot)$ then
   \begin{equation}\label{QCMI}
  \Delta_{\Phi}(\rho,\sigma)=I(A\!:\!BC)_{\rho}-I(A\!:\!C)_{\rho}=I(A\!:\!B|C)_{\rho}
  \end{equation}
   -- the quantum conditional mutual information of the state $\rho$ \cite{H-SCI,Wilde};

\item if $\rho$ and $\sigma$ are states in $\S(\H_{AB})$ and $\Phi=\Tr_A(\cdot)$ then
   $$
   \Delta_{\Phi}(\rho,\sigma)=D(\rho\shs\|\shs\sigma)-D(\rho_B\shs\|\shs\sigma_B)=D_A(\rho\shs\|\shs\sigma)
   $$
   -- the quantum conditional relative entropy of the states $\rho$ and $\sigma$ \cite{CondRE};

\item if $\rho=\sum_i p_i\rho_i\otimes |i\rangle\langle i|$ is a q-c state in $\S(\H_{AE})$ determined by any ensemble $\{p_i,\rho_i\}$ of quantum states in $\S(\H_A)$
   and a basis $\{|i\rangle\}$ in $\H_E$, $\sigma=\rho_A\otimes\rho_E$ and $\Phi$ is an arbitrary channel from $A$ to $B$ then\footnote{$\chi(\{p_i,\rho_i\})$ denotes the Holevo quantity of an ensemble $\{p_i,\rho_i\}$ defined in (\ref{chi-def}).}
   $$
   \Delta_{\Phi\otimes\id_E}(\rho,\sigma)=\chi(\{p_i,\rho_i\})-\chi(\{p_i,\Phi(\rho_i\}))
   $$
   -- the entropic disturbance of  $\{p_i,\rho_i\}$ by the channel $\Phi$ \cite{ED-1,ED-2,H&Sh-ED};

\item if $\rho$ is a pure state in $\S(\H_{AR})$, $\sigma=\rho_A\otimes\rho_{R}$ and $\Phi$ is an arbitrary channel from $A$ to $B$ then
   $$
   \Delta_{\Phi\otimes\id_R}(\rho,\sigma)=2S(\rho_A)-I(B\!:\!R)_{\Phi\otimes\id_R(\rho)}=I(\widehat{\Phi},\rho_A)
   $$
   -- the mutual information of any channel $\widehat{\Phi}$  complementary to the channel $\Phi$ at the state $\rho_A$ \cite{H-SCI,H-comp-ch};\footnote{Since
  a complementary channel $\widehat{\Phi}$ is defined up to the isometrical equivalence, the mutual information $I(\widehat{\Phi},\rho_A)$ is uniquely determined by $\Phi$ and equal to $2S(\rho_A)-I(\Phi,\rho_A)$ \cite{H-comp-ch}.}

\item if $\rho$ and $\sigma$ are arbitrary states in $\S(\H_A)$ and $\Phi(\varrho)=\sum_i P_i\varrho P_i$
   is a pinching channel determined by any set $\{P_i\}$ of mutually orthogonal projectors such that $[P_i\sigma]=0$ for all $i$ and $\,\sum_iP_i=I_{A}$ then
   $$
   \Delta_{\Phi}(\rho,\sigma)=S(\Phi(\rho))-S(\rho)
   $$
   -- the entropy gain  of the channel $\Phi$ at the state $\rho$;

\item if $\rho$ is a state in $\S(\H_{AB})$, $\sigma=\rho_A\otimes\rho_{B}$ and $\Phi(\varrho)=\sum_i[\Tr_B (I_A\otimes M_i)\varrho\shs]\otimes|i\rangle\langle i|$, where
$\{M_i\}$ is a POVM on $\H_B$ and $\{|i\rangle\}$ is a basic in $\H_E$,
then
   $$
   \Delta_{\Phi}(\rho,\sigma)=I(A\!:\!B)_{\rho}-I(A\!:\!E)_{\Phi(\rho)}=D_B^{\M}(\rho)
   $$
   -- the unoptimised quantum discord of the state $\rho$ corresponding to the POVM $\M=\{M_i\}$ \cite{NQD,NQD+,Wilde};

  \item if $\Phi:A\to B$ is a degradable channel (cf.\cite{D-Ch}) and $\Theta:B\to E$ is a channel such that $\Theta\circ\Phi$ is a channel complementary to $\Phi$ then
 $$
\textstyle\frac{1}{2} \Delta_{\Theta\otimes\id_R}(\Phi\otimes I_R(\bar{\rho}),\Phi(\rho)\otimes\bar{\rho}_R)=I_c(\Phi,\rho)
 $$
 -- the coherent information of the channel $\Phi$ at a state $\rho\in\S(\H_A)$, where $\bar{\rho}$ is a given purification
 in $\S(\H_{AR})$ of the state $\rho$ \cite{H-SCI,Wilde}.
 \end{itemize}

In addition to the above examples one should say that in many basic inequalities in quantum information theory
the gap (the difference between the r.h.s. and the l.h.s.) can be expressed via the function $\Delta_{\Phi}(\rho,\sigma)$. This holds, for example,
for the first and the second chain rules for the mutual information of a quantum channel $I(\Phi,\rho)$ (defined in Section 2.1). Indeed, it is easy to see that
for any channels $\Phi:A\to B$ and $\Psi:B\to C$ and an arbitrary state $\rho$ in $\S(\H_A)$ the following expressions hold
$$
I(\Phi,\rho)-I(\Psi\circ\Phi,\rho)=\Delta_{\Psi\otimes\id_R}(\omega_{BR},\omega_{B}\otimes\omega_{R})
$$
and
\begin{equation}\label{chain-2}
I(\Psi,\Phi(\rho))-I(\Psi\circ\Phi,\rho)=\Delta_{\Tr_E(\cdot)}(\Psi\otimes\id_{RE}(\omega),\Psi(\omega_{B})\otimes\omega_{RE}),
\end{equation}
where $\,\omega=V_{\Phi}\otimes I_R\cdot \bar{\rho}\shs\cdot V_{\Phi}^*\otimes I_R\,$ is a state in $\S(\H_{BER})$ defined by means of the Stinespring representation (\ref{St-rep}) of the channel $\Phi$ and a
given purification $\bar{\rho}\in\S(\H_{AR})$ of the state $\rho$.

For many of the above "special realizations" of $\Delta_{\Phi}(\rho,\sigma)$  it is known that the corresponding characteristic is a lower semicontinuous function of a state $\rho$. In the first and the third  cases
this was proved in \cite{CMI} and \cite{H&Sh-ED} correspondingly. In the fourth and fifth cases this follows from the lower semicontinuity of QRE, since
$I(\widehat{\Phi},\rho_A)=D(\widehat{\Phi}\otimes\id_R(\rho)\|\shs\widehat{\Phi}(\rho_A)\otimes \rho_R)\,$ and $\,S(\Phi(\rho))-S(\rho)=D(\rho\shs\|\Phi(\rho))\,$ for any pinching channel $\Phi$.\smallskip

It turns out that the lower semicontinuity of all the characteristics expressed via the function $\Delta_{\Phi}(\rho,\sigma)$ is a corollary of one general result
presented in the first part of the following theorem.
\smallskip

\begin{theorem}\label{main} A) \emph{For an arbitrary  quantum operation $\Phi:\T(\H_A)\to\T(\H_B)$ the function
$\,\Delta_{\Phi}(\rho,\sigma)=D(\rho\shs\|\shs\sigma)-D(\Phi(\rho)\|\shs\Phi(\sigma))$
is lower semicontinuous on the set}
$$
\left\{(\rho,\sigma)\in \T_+(\H_A)\times\T_+(\H_A)\,|\,D(\Phi(\rho)\|\shs\Phi(\sigma))<+\infty\right\}.
$$

B) \emph{The function $(\rho,\sigma,\Phi)\mapsto\Delta_{\Phi}(\rho,\sigma)$
is lower semicontinuous on the set
$$
\left\{(\rho,\sigma,\Phi)\in \T_+(\H_A)\times\T_+(\H_A)\times\F_{\leq1}(A,B)\,|\,D(\Phi(\rho)\|\shs\Phi(\sigma))<+\infty\right\},
$$
where $\F_{\leq1}(A,B)$ is the set of all quantum operations from $A$ to $B$ equipped with the strong convergence topology (see Section 2.1).}
\end{theorem}
\smallskip

\emph{Proof.} A) Assume first that $\Phi$ is a quantum channel with the Stinespring representation
$\Phi(\varrho)=\mathrm{Tr}_E V\varrho V^*$, $\varrho\in\T(\H_A)$, where $V$ is an isometry from $\,\mathcal{H}_{A}$ to $\mathcal{H}_{BE}$.

Let $\,\{\rho_n\}$ and $\{\sigma_n\}$ be sequences of operators in $\,\T_+(\H_A)$ converging, respectively,
to operators  $\rho_0$ and $\sigma_0$ such that $D(\Phi(\rho_n)\|\shs\Phi(\sigma_n))<+\infty$ for all $n\geq0$. We have to show
that
\begin{equation}\label{D-cont}
\liminf_{n\to+\infty}\Delta_{\Phi}(\rho_n,\sigma_n)\geq\Delta_{\Phi}(\rho_0,\sigma_0).
\end{equation}

If $\sigma_0=0$ then $\rho_0=0$ (otherwise $D(\Phi(\rho_0)\|\shs\Phi(\sigma_0))=+\infty$) and
(\ref{D-cont}) follows directly from the
monotonicity property (\ref{m-prop}). So, we will assume that $\sigma_0\neq0$ and, hence, $\Phi(\sigma_0)\neq0$ (as $\Phi$ is a channel).\smallskip

Since $V$ is an isometry, we have
\begin{equation*}%\label{a-n}
a_n\doteq D(V\rho_nV^*\|\shs V\sigma_nV^*)=D(\rho_n\|\shs\sigma_n),\;\, \forall n\geq0.
\end{equation*}
Let $\{P^n_m\}_{n\geq0,m\geq m_0}$ be a double sequence of finite rank projectors in $\B(\H_B)$
completely consistent  with the sequence $\{\Phi(\sigma_n)\}$ (Definition \ref{scs-def} in Section 2.1) which exists by Lemma 4 in \cite{REC}. Consider the double sequences
$$
a_n^m=D((P^n_m\otimes I_E)V\rho_nV^*(P^n_m\otimes I_E)\|\shs (P^n_m\otimes I_E)V\sigma_nV^*(P^n_m\otimes I_E))\leq+\infty,
$$
$$
b_n^m=D((\bar{P}^n_m\otimes I_E)V\rho_nV^*(\bar{P}^n_m\otimes I_E)\|\shs (\bar{P}^n_m\otimes I_E)V\sigma_nV^*(\bar{P}^n_m\otimes I_E))\leq+\infty
$$
and
$$
\hat{a}_n^m=D(P^n_m\Phi(\rho_n)P^n_m\|\shs P^n_m\Phi(\sigma_n))<+\infty,
$$
$$
\hat{b}_n^m=D(\bar{P}^n_m\Phi(\rho_n)\bar{P}^n_m\|\shs \bar{P}^n_m\Phi(\sigma_n))<+\infty,
$$
$ n\geq0,m\geq m_0$, where $\bar{P}^n_m=I_B-P^n_m$. By Lemma 4 in \cite{L-2} we have
\begin{equation}\label{L-imp}
  a_n\geq a_n^m+b_n^m.
\end{equation}
Lemma \ref{re-l} below implies that
\begin{equation}\label{K-imp}
\hat{a}_n\doteq D(\Phi(\rho_n)\|\shs\Phi(\sigma_n))=\hat{a}_n^m+\hat{b}_n^m+D(\Phi(\rho_n)\shs\|\textstyle\frac{1}{2}(\Phi(\rho_n)+U^n_m\Phi(\rho_n)[U^n_m]^*))
\end{equation}
for all $n\geq0$ and $m\geq m_0$, where $U^n_m=2P^n_m-I_B$ is a unitary operator.

Since $\,\hat{b}_n^m\leq b_n^m$ by the monotonicity  property (\ref{m-prop}) of the QRE, it follows from (\ref{L-imp}) and (\ref{K-imp}) that
\begin{equation}\label{B-ineq}
c_n^m\doteq a_n^m-\hat{a}_n^m-D(\Phi(\rho_n)\shs\|\textstyle\frac{1}{2}(\Phi(\rho_n)+U^n_m\Phi(\rho_n)[U^n_m]^*))\leq a_n-\hat{a}_n=\Delta_{\Phi}(\rho_n,\sigma_n)
\end{equation}
for all $n\geq0$ and $m\geq m_0$.

By Lemma \ref{b-lemma} below the properties of the double sequence $\{P^n_m\}_{n\geq0,m\geq m_0}$ imply that
\begin{equation}\label{est-b}
\lim_{n\to+\infty}\hat{a}_n^m=\hat{a}_0^m<+\infty\quad\forall m\geq m_0.
\end{equation}

Since $P^n_m$ tends to $P^0_m$ in the operator norm topology as $n\to+\infty$ and $\Phi(\rho_n)\leq \Phi(\rho_n)+U^n_m\Phi(\rho_n)[U^n_m]^*$  for all $n\geq0$ and $m\geq m_0$, by using Proposition 2 in \cite{DTL} we obtain
\begin{equation}\label{est-c}
\lim_{n\to+\infty}D(\Phi(\rho_n)\shs\|\textstyle\frac{1}{2}(\Phi(\rho_n)+U^n_m\Phi(\rho_n)[U^n_m]^*))=
D(\Phi(\rho_0)\shs\|\textstyle\frac{1}{2}(\Phi(\rho_0)+U^0_m\Phi(\rho_0)[U^0_m]^*))<+\infty.
\end{equation}

Since $\,(P^n_m\otimes I_E)V\omega_nV^*(P^n_m\otimes I_E)\,$ tends to $\,(P^0_m\otimes I_E)V\omega_0V^*(P^0_m\otimes I_E)$ as $n\to+\infty$ for all $m\geq m_0$, $\omega=\rho,\sigma$, the lower semicontinuity of the quantum relative entropy implies that
\begin{equation*}%\label{est-a}
\liminf_{n\to+\infty}a_n^m\geq a_0^m
\end{equation*}
for all $m\geq m_0$. This and the limit relations (\ref{est-b}) and (\ref{est-c}) show that
\begin{equation*}%\label{est-a}
\liminf_{n\to+\infty}c_n^m\geq c_0^m
\end{equation*}
for all $m\geq m_0$. So, it follows from (\ref{B-ineq}) that to prove  (\ref{D-cont})
it suffices to show that
\begin{equation*}%\label{est-a}
\lim_{m\to+\infty}c_n^m=\Delta_{\Phi}(\rho_n,\sigma_n)\leq+\infty
\end{equation*}
for all $n\geq0$. This can be done by noting that
\begin{equation}\label{est-a}
\lim_{m\to+\infty}D(\Phi(\rho_n)\shs\|\textstyle\frac{1}{2}(\Phi(\rho_n)+U^n_m\Phi(\rho_n)[U^n_m]^*))=0
\end{equation}
for all $n\geq0$, as Lemma 4 in \cite{L-2} implies that
$$
\lim_{m\to+\infty}\hat{a}_n^m=\hat{a}_n<+\infty\quad \textrm{and} \quad \lim_{m\to+\infty}a_n^m=a_n\leq+\infty.
$$

Since the condition $D(\Phi(\rho_n)\|\shs\Phi(\sigma_n))<+\infty$ implies that $\supp\Phi(\rho_n)\subseteq\supp\Phi(\sigma_n)$ for all $n\geq0$, the properties of the double sequence $\{P^n_m\}_{n\geq0,m\geq m_0}$ show that $U^n_m\Phi(\rho_n)[U^n_m]^*$ tends to $\Phi(\rho_n)$ as $m\to+\infty$ for each $n\geq0$. Hence,
the limit relation (\ref{est-a}) is proved easily by using Proposition 2 in \cite{DTL}.\smallskip

The validity of claim A in the case when $\Phi$ is an arbitrary quantum operation follows from claim B proved below (by using
the part of claim A proved above).\smallskip

B) Let $\,\{\rho_n\}$ and $\{\sigma_n\}$ be sequences of operators in $\,\T_+(\H_A)$ converging, respectively,
to operators  $\rho_0$ and $\sigma_0$. Let $\,\{\Phi_n\}$ be a sequence of quantum operations in $\,\F_{\leq1}(A,B)$ strongly converging to a
quantum operation $\Phi_0$  such that $D(\Phi_n(\rho_n)\|\shs\Phi_n(\sigma_n))<+\infty$ for all $n\geq0$. We have to show
that
\begin{equation}\label{D-cont++}
\liminf_{n\to+\infty}\Delta_{\Phi_n}(\rho_n,\sigma_n)\geq\Delta_{\Phi_0}(\rho_0,\sigma_0).
\end{equation}

Assume first that $\,\{\Phi_n\}$ is a sequence of quantum channels strongly converging to a
quantum channel $\Phi_0$. By Theorem 7 in \cite{CSR} there exist a system $E$ and a sequence $\{V_n\}$  of isometries  from $\,\mathcal{H}_{A}$ into $\mathcal{H}_{BE}$ strongly converging to an isometry $V_0$ such that $\mathrm{\Phi}_n(\varrho)=\mathrm{Tr}_E V_n\varrho V^*_n$ for all $\,n\geq0$.

It is clear that the operators
$\varrho_n=V_n\rho_nV_n^*$ and $\varsigma_n=V_n\sigma_nV_n^*$ in $\T_+(\H_{BE})$ tend, respectively, to the operators
$\varrho_0=V_0\rho_0V_0^*$ and $\varsigma_0=V_0\sigma_0V_0^*$ as $n\to+\infty$. Since
all the operators  $V_n$ are isometries, we have
\begin{equation*}%\label{V-out-l-r+}
D(\varrho_n\|\shs \varsigma_n)=D(\rho_n\|\shs \sigma_n)\quad \forall n\geq0.
\end{equation*}
Hence,  $\Delta_{\Phi_n}(\rho_n,\sigma_n)=\Delta_{\Theta}(\varrho_n,\varsigma_n)$ for all $n\geq0$, where
$\Theta=\Tr_E(\cdot)$ is a channel from $BE$ to $B$.  Thus, limit relation (\ref{D-cont++}) follows from the proved part of claim  A.\smallskip

Assume now  that $\,\{\Phi_n\}$ is a sequence of quantum operations  strongly converging to a
quantum operation $\Phi_0$. Lemma 2 in \cite{REC} and its proof show the existence of a system $C$ and a sequence $\{\widetilde{\Phi}_n\}$ of quantum channels  from $\T(\H_A)$ to $\T(\H_B\oplus\H_C)$ strongly converging to a quantum channel $\widetilde{\Phi}_0$ such that
\begin{equation*}%\label{phi-n-rep}
\Phi_n(\varrho)=P_B\widetilde{\Phi}_n(\varrho)=\widetilde{\Phi}_n(\varrho)P_B\quad \forall \varrho\in\T(\H_A),\; \forall n\geq0,
\end{equation*}
where $P_B$ is the projector onto the subspace $\H_B$ of $\mathcal{H}_{B}\oplus\mathcal{H}_{C}$.
By identity (\ref{D-sum}) we have
\begin{equation}\label{D-rep}
D(\widetilde{\Phi}_n(\rho_n)\|\shs \widetilde{\Phi}_n(\sigma_n))=D(\Phi_n(\rho_n)\|\shs \Phi_n(\sigma_n))+D(\Psi_n(\rho_n)\|\shs \Psi_n(\sigma_n)),\quad \forall n\geq0,
\end{equation}
where $\Psi_n(\varrho)=P_C\widetilde{\Phi}_n(\varrho)$, $P_C$ is the projector onto the subspace $\H_C$ of $\mathcal{H}_{B}\oplus\mathcal{H}_{C}$.

Since we do not assume that $D(\rho_n\|\shs \sigma_n)$ is finite for all $n\geq0$, we can not guarantee that $D(\widetilde{\Phi}_n(\rho_n)\|\shs \widetilde{\Phi}_n(\sigma_n))$
is finite. So, we can not directly apply the above part of the proof to the sequence  $\{\widetilde{\Phi}_n\}$. Nevertheless, we may reduce attention to the case
$$
\liminf_{n\to+\infty}D(\widetilde{\Phi}_n(\rho_n)\|\shs \widetilde{\Phi}_n(\sigma_n))\leq\liminf_{n\to+\infty}D(\rho_n\|\shs \sigma_n)<+\infty,
$$
since otherwise the limit relation  (\ref{D-cont++}) holds trivially. Thus, by passing to a subsequence we may assume
that $D(\widetilde{\Phi}_n(\rho_n)\|\shs \widetilde{\Phi}_n(\sigma_n))<+\infty$ for all $n>0$.

If $D(\widetilde{\Phi}_0(\rho_0)\|\shs \widetilde{\Phi}_0(\sigma_0))<+\infty$ then by applying the above part of the proof to the sequence  $\{\widetilde{\Phi}_n\}$
of quantum channels we obtain
\begin{equation*}%\label{D-cont+++}
\liminf_{n\to+\infty}\Delta_{\widetilde{\Phi}_n}(\rho_n,\sigma_n)\geq\Delta_{\widetilde{\Phi}_0}(\rho_0,\sigma_0).
\end{equation*}
This and (\ref{D-rep}) imply (\ref{D-cont++}), since
\begin{equation}\label{D-rel}
\liminf_{n\to+\infty}D(\Psi_n(\rho_n)\|\shs \Psi_n(\sigma_n))\geq D(\Psi_0(\rho_0)\|\shs \Psi_0(\sigma_0))
\end{equation}
by the lower semicontinuity of the QRE.

If $D(\widetilde{\Phi}_0(\rho_0)\|\shs \widetilde{\Phi}_0(\sigma_0))=+\infty$ then expression (\ref{D-rep}) and the assumed finiteness of  $D(\Phi_0(\rho_0)\|\shs \Phi_0(\sigma_0))$ show that $D(\Psi_0(\rho_0)\|\shs \Psi_0(\sigma_0))=+\infty$. So, it follows from (\ref{D-rel}) that
$$
\lim_{n\to+\infty}D(\Psi_n(\rho_n)\|\shs \Psi_n(\sigma_n))=+\infty.
$$
This implies
$$
\lim_{n\to+\infty}\Delta_{\Phi_n}(\rho_n,\sigma_n)=+\infty,
$$
since $\Delta_{\Phi_n}(\rho_n,\sigma_n)\geq D(\Psi_n(\rho_n)\|\shs \Psi(\sigma_n))$ for all $n$ due to (\ref{D-rep}).
Hence, the limit relation  (\ref{D-cont++}) holds trivially in this case. $\Box$

\smallskip

\begin{lemma}\label{re-l} \emph{Let $\rho$ and $\sigma$ be operators in $\T_+(\H)$ and $P$
a projector in $\B(\H)$ such that $P\sigma=\sigma P$. Then
\begin{equation}\label{re-l+}
D(\rho\shs\|\shs\sigma)=\textstyle D(P\rho P\shs\|\shs P\sigma)+D(\bar{P}\rho \bar{P}\shs\|\shs\bar{P}\sigma)+
D(\rho\shs\|\shs\frac{1}{2}(\rho+U\rho U^*)),
\end{equation}
where $U=2P-I_{\H}$ is a unitary operator.}
\end{lemma}

\emph{Proof.} We may assume that the operators $\rho$ and $\sigma$ are nonzero, since otherwise (\ref{re-l+}) holds trivially.

Since $U\sigma U^*=\sigma$ and $U^*=U$, by using Donald's identity (\ref{Donald}) we obtain
\begin{equation*}%\label{est-g}
\begin{array}{rl}
D(\rho\shs\|\shs \sigma)\,=&\!\! \textstyle\frac{1}{2}D(\rho\shs\|\shs \sigma)+\frac{1}{2}D(U\rho U^*\|\shs \sigma)=D(\textstyle\frac{1}{2}(\rho+U\rho U^*)\|\shs \sigma)\\\\+&\!\!\textstyle\frac{1}{2}D(\rho\shs\|\frac{1}{2}(\rho+U\rho U^*))
+\frac{1}{2}D(U\rho U^*\shs\|\frac{1}{2}(\rho+U\rho U^*))\\\\
=&\!\!D(P\rho P+\bar{P}\rho\bar{P}\|\shs \sigma)+D(\rho\shs\|\frac{1}{2}(\rho+U\rho U^*))\\\\
=&\!\!D(P\rho P\|\shs P\sigma)+D(\bar{P}\rho\bar{P}\|\shs \bar{P}\sigma)+D(\rho\shs\|\frac{1}{2}(\rho+U\rho U^*)),
\end{array}
\end{equation*}
where the last equality follows from identity (\ref{D-sum}). $\Box$\smallskip

\begin{lemma}\label{b-lemma} \emph{Let $\,\{\rho_n\}$ and $\{\sigma_n\}$ be sequences of operators in $\,\T_+(\H)$ converging, respectively,
to operators  $\rho_0$ and $\sigma_0\neq0$. If $\{P^n_m\}_{n\geq0,m\geq m_0}$ is a double sequence  of finite rank  projectors  completely consistent
with the sequence $\{\sigma_n\}$ (Definition 1 in Section 2.1) then}
\begin{equation}\label{D-cont-d+}
\lim_{n\to+\infty}\textstyle D(P^n_m\rho_nP^n_m\shs\|P^n_m\sigma_n)=D(P^0_m\rho_0P^0_m\shs\|P^0_m\sigma_0)<+\infty\quad \forall m\geq m_0.
\end{equation}
\end{lemma}

\emph{Proof.} Since $P^n_m\rho_n P^n_m$ tends to $P^0_m\rho_0 P^0_m$ as $\,n\to+\infty\,$ and $\,\sup_{n\geq0}\rank P^n_m\rho_n P^n_m< +\infty$  by the  conditions in (\ref{P-prop}) and (\ref{P-prop+}), by using representation (\ref{re-exp}) we see that to prove (\ref{D-cont-d+}) it suffices to show that
\begin{equation}\label{D-cont-d++}
\lim_{n\to+\infty}\|P^n_m \ln (P^n_m\sigma_n)-P^0_m \ln (P^0_m\sigma_0)\|=0
\end{equation}
for any given $m\geq m_0$. By the second condition in (\ref{P-prop+}) we have $\rank P^n_m\sigma_n=\rank P^n_m$ for
all $n\geq0$. Hence the sequence  $\{P^n_m\sigma_n+\bar{P}^n_m\}_n$
consists of bounded nondegenerate operators and converges to the nondegenerate operator $P^0_m\sigma_0+\bar{P}^0_m$
in the operator  norm by the third condition in (\ref{P-prop+}). It follows that $P^n_m\sigma_n+\bar{P}^n_m\geq\epsilon I_{\H}$ for
all $n\geq0$ and some $\epsilon>0$. So, by Proposition VIII.20 in \cite{R&S} the sequence  $\{\ln (P^n_m\sigma_n+\bar{P}^n_m)\}_n$
converges to the operator $\ln (P^0_m\sigma_0+\bar{P}^0_m)$ in the operator  norm. This implies (\ref{D-cont-d++}), since
$$
P^n_m \ln (P^n_m\sigma_n)=P^n_m\ln (P^n_m\sigma_n+\bar{P}^n_m)\quad \forall n\geq0.\;\Box
$$

\subsection{Non-increase of the QRE local discontinuity jumps under action of quantum operations}

The lower semicontinuity of the relative entropy disturbance by quantum operations implies that the local discontinuity jumps of the QRE
do not increase under action of quantum  operations.

\begin{property}\label{NMP} \emph{Let $\,\{\rho_n\}$ and $\{\sigma_n\}$ be sequences of operators in $\,\T_+(\H_A)$ converging, respectively,
to operators  $\rho_0$ and $\sigma_0$ such that $D(\rho_0\|\shs\sigma_0)<+\infty$.}

A) \emph{If $\,\Phi$ is an arbitrary quantum operation from $A$ to $B$ then}\smallskip
\begin{equation}\label{D-j++}
\!\limsup_{n\to+\infty}D(\Phi(\rho_n)\|\shs\Phi(\sigma_n))-D(\Phi(\rho_0)\|\shs\Phi(\sigma_0))\leq\limsup_{n\to+\infty}D(\rho_n\|\shs\sigma_n)-D(\rho_0\|\shs\sigma_0).
\end{equation}

B) \emph{If $\,\{\Phi_n\}$ is a sequence of quantum operations from $A$ to $B$ strongly converging to a
quantum operation $\Phi_0$  then}
\begin{equation}\label{D-j+}
\!\limsup_{n\to+\infty}D(\Phi_n(\rho_n)\|\shs\Phi_n(\sigma_n))-D(\Phi_0(\rho_0)\|\shs\Phi_0(\sigma_0))\leq\limsup_{n\to+\infty}D(\rho_n\|\shs\sigma_n)-D(\rho_0\|\shs\sigma_0)
\end{equation}
\end{property}

\textbf{Note:} The l.h.s. of (\ref{D-j++}) and (\ref{D-j+}) are well defined due to the finiteness of  $D(\rho_0\|\shs\sigma_0)$ and the monotonicity of the QRE.
\medskip

By the lower semicontinuity of the QRE the quantities in the r.h.s. and the l.h.s. of (\ref{D-j++}) characterizes the
discontinuity of the functions $(\rho,\sigma)\mapsto D(\rho\shs\|\shs\sigma)$ and $(\rho,\sigma)\mapsto D(\Phi(\rho)\|\shs\Phi(\sigma))$
for given converging sequences $\{\rho_n\}$ and $\{\sigma_n\}$. So, the claim of Proposition \ref{NMP}A can be interpreted as a "contraction" of possible discontinuities of the quantum relative entropy by quantum operations.\medskip

\emph{Proof.} Since the QRE is a lower semicontinuous function of its arguments, claims A and B of the proposition follow, respectively, from claims A and B of Theorem \ref{main}
by Lemma \ref{vsl} in Section 2.1. $\Box$\smallskip

Proposition \ref{NMP} implies the following observations which can be treated as preserving of local continuity
of the QRE by a single quantum operation and by strongly converging sequence of quantum operations. It was originally proved
in \cite{REC} by completely different method (based on using the convergence criterion for the QRE proposed therein).\smallskip

\begin{corollary}\label{NMP-c} \emph{Let $\,\{\rho_n\}$ and $\{\sigma_n\}$ be sequences of operators in $\,\T_+(\H_A)$ converging, respectively,
to operators  $\rho_0$ and $\sigma_0$ such that}
\begin{equation}\label{D-c}
\lim_{n\to+\infty}D(\rho_n\|\shs\sigma_n)=D(\rho_0\|\shs\sigma_0)<+\infty.
\end{equation}

A) \emph{If $\,\Phi$ is an arbitrary quantum operation from $A$ to $B$ then}\smallskip
\begin{equation}\label{D-c+}
\lim_{n\to+\infty}D(\Phi(\rho_n)\|\shs\Phi(\sigma_n))=D(\Phi(\rho_0)\|\shs\Phi(\sigma_0))<+\infty.
\end{equation}

B) \emph{If $\,\{\Phi_n\}$ is a sequence of quantum operations from $A$ to $B$ strongly converging to a
quantum operation $\Phi_0$ then}
\begin{equation*}%\label{D-c++}
\lim_{n\to+\infty}D(\Phi_n(\rho_n)\|\shs\Phi_n(\sigma_n))=D(\Phi_0(\rho_0)\|\shs\Phi_0(\sigma_0))<+\infty.
\end{equation*}
\end{corollary}

\begin{remark}\label{NMP-c} Roughly speaking, the QRE of two states $\rho$ and $\sigma$ can be expressed as
$$
D(\rho\shs\|\shs\sigma)=F(\rho,\sigma)-S(\rho),\quad \textrm{where}\quad  F(\rho,\sigma)=\Tr\rho\shs(-\ln \sigma),
$$
i.e. as the difference between the nonnegative lower semicontinuous functions  $F$ and $S$. It
is easy to see that in general the local continuity of the functions  $F$ and $S$ is not preserved by quantum channels,
i.e. the validity of the limit relations
\begin{equation}\label{2-l-r}
\lim_{n\to+\infty}F(\rho_n,\sigma_n)=F(\rho_0,\sigma_0)<+\infty\quad \textrm{and} \quad
\lim_{n\to+\infty}S(\rho_n)=S(\rho_0)<+\infty
\end{equation}
for some sequences $\,\{\rho_n\}$ and $\{\sigma_n\}$ of states  in $\,\S_+(\H_A)$ converging
to states  $\rho_0$ and $\sigma_0$ \emph{do not imply}, respectively,
the validity of the limit relations
\begin{equation}\label{2-l-r+}
\lim_{n\to+\infty}F(\Phi(\rho_n),\Phi(\sigma_n))=F(\Phi(\rho_0),\Phi(\sigma_0))<+\infty
\end{equation}
and
\begin{equation}\label{2-l-r++}
\lim_{n\to+\infty}S(\Phi(\rho_n))=S(\Phi(\rho_0))<+\infty
\end{equation}
for any channel $\Phi$.

Thus, if $\,\{\rho_n\}$ and $\{\sigma_n\}$ are any converging sequences of states
such that the limit relations (\ref{2-l-r}) hold, but the limit relation (\ref{2-l-r++})
is not valid, then Corollary \ref{NMP-c}A states that the limit relation (\ref{2-l-r+})
is not valid as well and that the discontinuity jump of the function $F$ \emph{compensates}
the discontinuity jump of the function $S$ in such a way that
$$
\lim_{n\to+\infty}(F-S)(\Phi(\rho_n),\Phi(\sigma_n))=(F-S)(\Phi(\rho_0),\Phi(\sigma_0))<+\infty.
$$

Note that the monotonicity property of QRE means a similar compensation of possible increasing of the von Neumann entropy
$S$ under action of a quantum channel. This compensation is necessary for the non-increasing of the
QRE -- the function $F-S$.
\end{remark}\smallskip

\begin{remark}\label{NMP-r}  At first glance, Corollary \ref{NMP-c} is the only contribution
of Proposition \ref{NMP} to the task of proving local continuity (convergence)
of the QRE and the related functions. In fact, the more general claim of Proposition \ref{NMP}
gives the possibility to prove local continuity of a given function $f$ via uniform approximation of this
function by appropriate functions $f_n$, for which a relation similar to (\ref{D-j++}) can be established
with the r.h.s. tending to zero as $n\to+\infty$. This approach is used in the proof of Proposition \ref{vn-fun}
in Section 4.1.6 below.
\end{remark}\medskip

There is a class of quantum operations $\Phi$ for which relation (\ref{2-l-r++}) holds provided that
the second limit relation in (\ref{2-l-r}) is valid. Quantum operations from this class are characterized by the following
equivalent properties \cite[Theorem 1]{PFE}:
\begin{itemize}
  \item $S(\Phi(\rho))<+\infty$ for any state $\rho$ such that $S(\rho)<+\infty$;
  \item $\sup\{S(\Phi(\rho))\,|\, \rho\in\S(\H), \rank\rho=1\}<+\infty$.
\end{itemize}
The first of these properties means that such operations \emph{preserve finiteness of the von Neumann entropy}, so they are
called PFE-operations in \cite{PFE}, where their  detailed classification can be found.\smallskip

\begin{corollary}\label{NMP-c+} \emph{Let $\,\{\rho_n\}$ and $\{\sigma_n\}$ be sequences of operators in $\,\T_+(\H_A)$ converging, respectively,
to operators  $\rho_0$ and $\sigma_0$ such that
\begin{equation}\label{T-c}
\lim_{n\to+\infty}\Tr \rho_n(-\ln\sigma_n)=\Tr \rho_0(-\ln\sigma_0)<+\infty.
\end{equation}
Then
\begin{equation}\label{T-c+}
\lim_{n\to+\infty}\Tr \Phi(\rho_n)(-\ln\Phi(\sigma_n))=\Tr \Phi(\rho_0)(-\ln\Phi(\sigma_0))<+\infty
\end{equation}
for any PFE-operation $\Phi$.}
\end{corollary}\smallskip

\emph{Proof.} By Lemma \ref{vsl} in Section 2.1 condition (\ref{T-c}) implies the validity
of (\ref{D-c}) and the second limit relation in (\ref{2-l-r}). Hence, (\ref{D-c+}) holds by Corollary \ref{NMP-c}A
and (\ref{2-l-r++}) is valid, since $\Phi$ is a PFE-operation.  It is clear that  (\ref{T-c+}) follows from
(\ref{D-c+}) and (\ref{2-l-r++}). $\Box$

\subsection{Lower semicontinuity of the modulus  of the QRE joint convexity}

\subsubsection{The case of convex mixtures}

The joint convexity of the QRE means that
\begin{equation*}%\label{j-c-re}
  D\!\left(\sum_ip_i\rho_i\right\|\left.\sum_ip_i\sigma_i\right)\leq \sum_i p_iD(\rho_i\|\sigma_i)
\end{equation*}
for any ensembles $\{p_i,\rho_i\}$ and $\{p_i,\sigma_i\}$ of quantum states in $\S(\H)$ with the same probability distributions \cite{L-2,H-SCI,Wilde}.

Denote by $\P(\H)$ the set of all discrete ensembles of  states in $\S(\H)$. We will say that a sequence of ensembles $\mu_n=\{p^n_i,\rho^n_i\}$
in $\P(\H)$ $D_0$\emph{-converges} to an ensemble $\mu_0=\{p^0_i,\rho^0_i\}$ if
$$
\lim_{n\to+\infty}p^n_i\rho^n_i=p^0_i\rho^0_i\quad \forall i.
$$

Theorem \ref{main}A implies the following\smallskip

\begin{property}\label{j-c-g-ls} \emph{The nonnegative function
\begin{equation*}%\label{j-c-re-g}
 (\{p_i,\rho_i\},\{p_i,\sigma_i\})\mapsto \sum_i p_iD(\rho_i\|\sigma_i)-D\!\left(\sum_ip_i\rho_i\shs\right\|\left.\sum_ip_i\sigma_i\right)
\end{equation*}
is lower semicontinuous on the set
\begin{equation}\label{P-set}
\left\{(\{p_i,\rho_i\},\{p_i,\sigma_i\})\in\P(\H)\times\P(\H)\left|\, D\!\left(\sum_ip_i\rho_i\shs\right\|\left.\sum_ip_i\sigma_i\right)<+\infty \right\}\right.
\end{equation}
with respect to the $D_0$-convergence in $\P(\H)\times\P(\H)$.}
\end{property}
\smallskip

\emph{Proof.} For a given pair of ensembles $\mu=\{p_i,\rho_i\}$ and $\nu=\{p_i,\sigma_i\}$ belonging to the set in (\ref{P-set}) introduce the q-c states
$$
\hat{\mu}=\sum_i p_i\rho_i\otimes|i\rangle\langle i|\quad \textrm{and}\quad \hat{\nu}=\sum_i p_i\sigma_i\otimes|i\rangle\langle i|
$$
in $\S(\H\otimes\H_E)$, where $\{|i\rangle\}$ is a basic in a separable Hilbert space $\H_E$. Then identity (\ref{D-sum})
implies that
$$
\sum_i p_iD(\rho_i\|\sigma_i)=D(\hat{\mu}\|\shs\hat{\nu}),
$$
while
$$
D\!\left(\sum_ip_i\rho_i\shs\right\|\left.\sum_ip_i\sigma_i\right)=D(\Phi(\hat{\mu})\|\Phi(\hat{\nu}))
$$
where $\Phi=\Tr_E(\cdot)$. Thus, the claim of the proposition follows directly from Theorem \ref{main}A. $\Box$
\smallskip

By using the lower semicontinuity of the QRE and Lemma \ref{D-A} in Section 2.1 it is easy to show that the function $(\{p_i,\rho_i\},\{p_i,\sigma_i\})\mapsto D\!\left(\sum_ip_i\rho_i\shs\right\|\left.\sum_ip_i\sigma_i\right)$
is lower semicontinuous on the set $\P(\H)\times\P(\H)$ with respect to the $D_0$-convergence. Thus,
Proposition \ref{j-c-g-ls} implies  (by Lemma \ref{vsl} in Section 2.1) the following convergence condition for the quantum relative entropy.
\smallskip

\begin{corollary}\label{j-c-g-c} \emph{Let $\{\{p^n_i,\rho^n_i\}\}_n$ and $\{\{p^n_i,\sigma^n_i\}\}_n$ be  sequences of ensembles in $\P(\H)$ $D_0$-converging to ensembles $\{p^0_i,\rho^0_i\}$ and $\{p^0_i,\sigma^0_i\}$ correspondingly. If
\begin{equation*}%\label{j-c-lr}
\lim_{n\to+\infty}\sum_i p^n_iD(\rho^n_i\|\sigma^n_i)=\sum_i p^0_iD(\rho^0_i\|\sigma^0_i)<+\infty
\end{equation*}
then}
\begin{equation*}%\label{j-c-lr}
\lim_{n\to+\infty}D\!\left(\sum_ip^n_i\rho^n_i\right\|\left.\sum_ip^n_i\sigma^n_i\right)=D\!\left(\sum_ip^0_i\rho^0_i\right\|\left.\sum_ip^0_i\sigma^0_i\right)<+\infty.
\end{equation*}
\end{corollary}
\smallskip

In Section 4.1.5 we will obtain a generalization of the claim of Corollary \ref{j-c-g-c}
to the case of arbitrary $D_0$-converging sequences of discrete ensembles.

\subsubsection{The case of finite and countable sums}

The joint convexity and lower semicontinuity of the QRE along with identity (\ref{D-mul}) imply that
\begin{equation}\label{JC++}
D\!\left.\left(\sum_i\rho_i\shs\right\|\sum_i\sigma_i\right)\leq \sum_i D(\rho_i\|\sigma_i)
\end{equation}
for any finite or countable sets $\{\rho_i\}$ and $\{\sigma_i\}$ of operators in $\T_+(\H)$ such that
$\sum_i\Tr\rho_i$ and $\sum_i\Tr\sigma_i$ are finite.

Denote by $\T^{\infty}_+(\H)$ the set $\,\{\{\varrho_i\}_{i=1}^{+\infty}\subset\T_+(\H)\,|\sum_i\Tr\varrho_i<+\infty\}\,$. We will say that a sequence
$\{\{\varrho^n_i\}_i\}_n\subset\T^{\infty}_+(\H)$ converges to $\{\varrho^0_i\}\in\T^{\infty}_+(\H)$ if
$\sum_i\varrho^n_i$ converges to $\sum_i\varrho^0_i$ in $\T(\H)$
or, equivalently,\footnote{This equivalence follows from Lemma \ref{D-A} in Section 2.1.}
$$
\lim_{n\to+\infty}\varrho^n_i=\varrho^0_i\;\;\forall i\quad \textrm{and}\quad \lim_{n\to+\infty}\sum_i\Tr\varrho^n_i=\sum_i\Tr\varrho^0_i .
$$

The same arguments as in the proof of Proposition \ref{j-c-g-ls} allows us to derive from Theorem \ref{main}A the following\smallskip

\begin{property}\label{c-sum} \emph{The function $\,(\{\rho_i\},\{\sigma_i\})\mapsto\sum_i D(\rho_i\|\sigma_i)-D(\sum_i\rho_i\shs\|\sum_i\sigma_i)\,$ is lower semicontinuous on the set}
$$
\left\{(\{\rho_i\},\{\sigma_i\})\in[\T^{\infty}_+(\H)]^{\times 2}\,\left|\, D\!\!\left.\left(\sum_i\rho_i\shs\right\|\sum_i\sigma_i\right)<+\infty\right.\right\}.
$$
\end{property}

This result and Lemma \ref{vsl} in Section 2.1 imply directly the following observation concerning
convergence of the QRE for  finite and  countable sums.\smallskip

\begin{corollary}\label{c-sum-c} \emph{Let $\,\{\{\rho^n_i\}_{n\geq 0}\}_{i\in I}$ and  $\{\{\sigma^n_i\}_{n\geq 0}\}_{i\in I}$  be finite or countable sets of converging sequences of operators in $\,\T_+(\H)$ such that
$\sum_{i\in I}D(\rho^0_i\|\sigma^0_i)<+\infty$,
\begin{equation*}%\label{K-conv}
  \lim_{n\to+\infty}\sum_{i\in I}\Tr\rho^n_i=\sum_{i\in I}\Tr\rho^0_i<+\infty\quad \textit{and} \quad \lim_{n\to+\infty}\sum_{i\in I}\Tr\sigma^n_i=\sum_{i\in I}\Tr\sigma^0_i<+\infty,
\end{equation*}
where $\,\rho^0_i=\lim_{n\to+\infty}\rho^n_i\,$ and $\,\sigma^0_i=\lim_{n\to+\infty}\sigma^n_i$, $i\in I$. Then}\footnote{The symbol "$\djn$" is defined in (\ref{dj}).}
\begin{equation}\label{g-r-sum}
\djn \left(\left\{D\!\!\left.\left(\sum_{i\in I}\rho^n_i\shs\right\|\sum_i\sigma^n_i\right)\right\}_{\!\!n}\right)\leq \djn \left(\left\{\sum_{i\in I} D(\rho^n_i\|\sigma^n_i)\right\}_{\!\!n}\right).
\end{equation}
\emph{In particular, if
\begin{equation}\label{c-11+}
\lim_{n\to+\infty}\sum_{i\in I} D(\rho^n_i\|\sigma^n_i)=\sum_{i\in I} D(\rho^0_i\|\sigma^0_i)<+\infty
\end{equation}
then}
\begin{equation*}%\label{c-22+}
\lim_{n\to+\infty}D\!\!\left.\left(\sum_{i\in I}\rho^n_i\shs\right\|\sum_i\sigma^n_i\right)=D\!\!\left.\left(\sum_{i\in I}\rho^0_i\shs\right\|\sum_i\sigma^0_i\right)<+\infty.
\end{equation*}
\end{corollary}

\textbf{Note:} The l.h.s. of (\ref{g-r-sum}) is well defined, since $D(\sum_{i\in I}\rho^0_i\|\sum_i\sigma^0_i)<+\infty\,$ due to the condition $\sum_{i\in I}D(\rho^0_i\|\sigma^0_i)<+\infty\,$ and inequality (\ref{JC++}).
\medskip

If the set $I$ is finite then condition (\ref{c-11+}) means that
\begin{equation}\label{c-11++}
\lim_{n\to+\infty}D(\rho^n_i\|\sigma^n_i)=D(\rho^0_i\|\sigma^0_i)<+\infty\quad \forall i\in I.
\end{equation}
So, in this case the last claim of Proposition \ref{c-sum-c} agrees with Corollary 3 in \cite{DTL}, which states preserving
the QRE convergence under finite summation.\smallskip

If the set $I$ is countable then (\ref{c-11++}) is necessary but not sufficient for (\ref{c-11+}). So, in this case
we may treat (\ref{c-11+}) as a sufficient condition for preserving
the QRE convergence under countable summation. It can be compared with the condition given by Corollary 4 in \cite{REC}. Using
inequality (\ref{JC++}) and Dini's lemma one can show that the former condition is slightly weaker than the latter. The advantage
of Corollary \ref{c-sum-c} (in this article) is the general relation (\ref{g-r-sum}), which gives additional abilities in
analysis of local continuity of the QRE (see the proof of Proposition \ref{vn-fun} in Section 4.1.6  based on the main claim of Corollary \ref{new-fun-c} in Section 4.1.5, which is proved by using (\ref{g-r-sum})).

\section{Applications}

\subsection{Local continuity analysis of the QRE and the related functions}

\subsubsection{Generalization of Lindblad's lemma}

The most known and widely used result concerning convergence (local continuity) of the QRE is
presented in Lemma 4 in \cite{L-2}. It states that
\begin{equation*}%\label{L-conv}
\lim_{n\to+\infty}D(P_n\rho P_n\|P_n\sigma P_n)=D(\rho\|\sigma)\leq+\infty
\end{equation*}
for any operators $\rho$ and $\sigma$ in $\T_+(\H)$, where $\{P_n\}$ is any nondecreasing sequence of projectors converging to the unit operator in
the strong operator topology. The results of Section 3 implies the following strengthened version of this claim.\smallskip

\begin{property}\label{L-cond} \emph{If $\rho$ and $\sigma$ are operators in $\T_+(\H)$ such that $D(\rho\shs\|\sigma)<+\infty$ then
\begin{equation}\label{L-conv+}
\lim_{n\to+\infty}D(A_n\rho A_n\|A_n\sigma A_n)=D(A_0\rho A_0\|A_0\sigma A_0)
\end{equation}
for any sequence $\{A_n\}\subset\B(\H)$  converging to an operator $A_0\in\B(\H)$ in
the strong operator topology. If $A_0=I_{\H}$ and $\|A_n\|\leq1$ for all $n$ then (\ref{L-conv+}) holds regardless on the condition $D(\rho\shs\|\sigma)<+\infty$. }
\end{property}

\emph{Proof.} The uniform boundedness principle (see \cite[Theorem III.9]{R&S}) implies that\break $\sup_n\|A_n\|<+\infty$. So, due to identity (\ref{D-mul}) we may assume
that $\|A_n\|\leq1$ for all $n\geq0$. Then $\Phi_n=A_n(\cdot)A_n$ is a sequence of quantum operations strongly converging to the quantum operation $\Phi_0=A_0(\cdot)A_0$.
Hence, limit relation (\ref{L-conv+}) follows from Corollary \ref{NMP-c}B in Section 3.2.

The validity of (\ref{L-conv+}) in the case $D(\rho\shs\|\sigma)=+\infty$ provided that $A_0=I_{\H}$ and $\|A_n\|\leq1$ for all $n$ follows
from the lower semicontinuity of the QRE, since $D(A_n\rho A_n\|A_n\sigma A_n)\leq D(\rho\shs\|\shs\sigma)$ by the monotonicity
of the QRE under the quantum operation $\Phi_n=A_n(\cdot)A_n$. $\Box$

\begin{remark}\label{L-cond-r} The condition $D(\rho\shs\|\sigma)<+\infty$ in the main claim of Proposition \ref{L-cond}
is essential. Indeed, let $\rho$ and $\sigma$ be operators diagonizable in a given basis $\{|i\rangle\}$ in $\H$ such that $D(\rho\shs\|\sigma)=+\infty$ and $A_n=\sum_{i>n}|i\rangle\langle i|$ for all $n\in \mathbb{N}$.  Then it is easy to see that the sequence $\{A_n\}$ strongly converges to $A_0=0$,
$$
D(A_n\rho A_n\|A_n\sigma A_n)=+\infty\quad\forall n,\quad\textrm{while}\quad D(A_0\rho A_0\|A_0\sigma A_0)=D(0\|0)=0.
$$
\end{remark}

\subsubsection{The function $\Phi\mapsto D(\Phi(\rho)\|\shs\Phi(\sigma))$}

By using Corollary \ref{NMP-c}B in Section 3.2  in the  case $\rho_n=\rho_0$ and $\sigma_n=\sigma_0$  we obtain  the following
result about properties of the function $\Phi\mapsto D(\Phi(\rho)\|\shs\Phi(\sigma))$.
\smallskip

\begin{property}\label{good-f} \emph{If $\rho$ and $\sigma$ are operators in $\T_+(\H_A)$ such that
$D(\rho\shs\|\shs\sigma)<+\infty$ then the function $\Phi\mapsto D(\Phi(\rho)\|\shs\Phi(\sigma))$
is continuous on the set of all quantum operations from $A$ to any quantum system $B$ equipped with the strong convergence topology.}
\end{property}
\smallskip

The usefulness of this assertion can be illustrated by the following
\smallskip
\begin{example}\label{e-3}
Let $\{\Phi_t\}_{t\in \mathbb{R}_+}$
be an arbitrary strongly continuous family of quantum channels (for instance, a quantum dynamical semigroups). Then Proposition \ref{good-f} implies that
the function
$$
t\mapsto D(\Phi_t(\rho)\|\shs\Phi_t(\sigma))
$$
are  continuous on $\mathbb{R}_+$ for any input states $\rho$ and $\sigma$  such that
$D(\rho\shs\|\shs\sigma)$ is finite.\smallskip

If $\{\Phi_t\}$ is a semigroup then the general properties of the QRE show that the above function is non-increasing and lower semicontinuous
on $\mathbb{R}_+$. This implies only that it is right-continuous on $\mathbb{R}_+$. Thus, the claim about continuity of this function is not
trivial even in this case.
\end{example}

\subsubsection{The function $(\rho,\sigma,\eta,\theta)\mapsto D(\rho+\eta\shs\|\shs\sigma)-D(\rho\shs\|\shs\sigma+\theta)$}

Inequalities (\ref{re-ineq}) and (\ref{re-2-ineq-conc}) show that
$$
D(\rho+\eta\shs\|\shs\sigma)\geq D(\rho\shs\|\shs\sigma+\theta)-\Tr\theta-\Tr\sigma
$$
for any operators $\rho,\sigma,\eta$ and $\theta$ in  $\T_+(\H)$. \smallskip

Theorem \ref{main}A implies the following\smallskip

\begin{property}\label{sf-1} \emph{The function $(\rho,\sigma,\eta,\theta)\mapsto D(\rho+\eta\shs\|\shs\sigma)-D(\rho\shs\|\shs\sigma+\theta)$
is lower semicontinuous on the set}
\begin{equation*}%\label{sf-1-set}
  \left.\left\{(\rho,\sigma,\eta,\theta)\in[\T_+(\H)]^{\times 4}\,\right|\,D(\rho\shs\|\shs\sigma+\theta)<+\infty\right\}.
\end{equation*}
\end{property}

\emph{Proof.} Let $\H_R$ be a two-dimensional Hilbert space. For given operators $\rho,\sigma$ and $\theta$ in $\T_+(\H)$ consider the operators
$$
\hat{\rho}=\rho\otimes |0\rangle\langle0|,\quad \hat{\sigma}=\sigma\otimes |0\rangle\langle0|+\theta\otimes |1\rangle\langle1|
$$
in $\T_+(\H\otimes\H_R)$, where $\{|0\rangle,|1\rangle\}$ is a basis in $\H_R$. Then identity (\ref{D-sum}) implies that
$$
D(\hat{\rho}\shs\|\shs\hat{\sigma})=D(\rho\shs\|\shs\sigma)+D(0\shs\|\shs\theta)=D(\rho\shs\|\shs\sigma)+\Tr\theta,
$$
while
$$
D(\Tr_R\hat{\rho}\shs\|\shs\Tr_R\hat{\sigma})=D(\rho\shs\|\shs\sigma+\theta).
$$
Hence
\begin{equation}\label{one}
  D(\rho\shs\|\shs\sigma)=D(\rho\shs\|\shs\sigma+\theta)+\Delta_{\Tr_R(\cdot)}(\hat{\rho},\hat{\sigma})-\Tr\theta.
\end{equation}

Using Donald's identity (\ref{Donald}) and identities (\ref{D-mul}) and (\ref{D-c-id}) we obtain
\begin{equation}\label{two}
  D(\rho+\eta\shs\|\shs\sigma)+f(\rho,\eta)=D(\rho\shs\|\shs\sigma)+D(\eta\shs\|\shs\sigma)
  +\ln2\Tr(\rho+\eta)-\Tr\sigma,
\end{equation}
where $f(\rho,\eta)=D(\rho\shs\|\shs\frac{1}{2}(\rho+\eta))+D(\eta\shs\|\shs\frac{1}{2}(\rho+\eta))$. By applying
Proposition 2 in \cite{DTL} it is easy to show that $f(\rho,\eta)$ is a continuous function on $[\T_+(\H)]^{\times2}$.
Identity (\ref{D-c-id}) and inequality (\ref{re-ineq}) show that $f(\rho,\eta)\leq\ln2\Tr(\rho+\eta)$.

It follows from (\ref{one}) and (\ref{two}) that
$$
D(\rho+\eta\shs\|\shs\sigma)-D(\rho\shs\|\shs\sigma+\theta)=D(\eta\shs\|\shs\sigma)+\Delta_{\Tr_R(\cdot)}(\hat{\rho},\hat{\sigma})
+[\ln2\Tr(\rho+\eta)-f(\rho,\eta)]-\Tr(\sigma+\theta)
$$
for any $\rho,\sigma,\eta$ and $\theta$ in $\T_+(\H)$ such that $D(\rho\shs\|\shs\sigma+\theta)<+\infty$. Thus, the claim
of the proposition follows from the lower semicontinuity of the QRE, Theorem \ref{main}A and continuity of the function
$f(\rho,\eta)$ mentioned above. $\Box$
\smallskip

Proposition \ref{sf-1}, the  lower semicontinuity of QRE and Lemma \ref{vsl} imply the following

\smallskip

\begin{corollary}\label{sf-1-c}
\emph{Let $\,\{\rho^1_n\}$, $\{\rho^2_n\}$, $\{\sigma^1_n\}$ and $\,\{\sigma^2_n\}$ be sequences of operators in $\T_+(\H)$ converging, respectively,
to operators  $\rho^1_0$, $\rho^2_0$, $\sigma^1_0$ and $\sigma^2_0$ such that $\,\rho^2_n\leq \rho^1_n$ and $\,\sigma^1_n\leq \sigma^2_n$ for all $\,n\geq0$. If $D(\rho^1_0\|\shs\sigma^1_0)<+\infty$ then $D(\rho^2_0\|\shs\sigma^2_0)<+\infty$ and}\footnote{The symbol "$\djn$" is defined in (\ref{dj}).}
$$
\djn(\{D(\rho^2_n\|\shs\sigma^2_n)\})\leq \djn(\{D(\rho^1_n\|\shs\sigma^1_n)\}).
$$
\emph{In particular, if
\begin{equation*}%\label{c-11+}
\lim_{n\to+\infty}D(\rho^1_n\|\shs\sigma^1_n)=D(\rho^1_0\|\shs\sigma^1_0)<+\infty
\end{equation*}
then}
\begin{equation*}%\label{c-22+}
\lim_{n\to+\infty}D(\rho^2_n\|\shs\sigma^2_n)=D(\rho^2_0\|\shs\sigma^2_0)<+\infty.
\end{equation*}
\end{corollary}\smallskip

The last claim of Corollary \ref{sf-1-c} coincides with the claim of Proposition 2 in \cite{DTL} proved by
rather indirect and complex way.

\subsubsection{The function $(\rho,\sigma,\eta,\theta)\mapsto \Tr (\rho+\eta)(-\ln\sigma)-\Tr \rho(-\ln(\sigma+\theta))$}

The operator monotonicity of the logarithm and the obvious inequality $\sigma\leq I_{\H}\Tr\sigma $ shows that
$$
\Tr (\rho+\eta)(-\ln\sigma)\geq \Tr \rho(-\ln(\sigma+\theta))-\Tr\eta\ln\Tr\sigma
$$
for any operators $\rho,\sigma,\eta$ and $\theta$ in  $\T_+(\H)$.\footnote{According to (\ref{H-fun}) we assume that $0\ln0=0$ and $\Tr\varrho\ln0=+\infty$ for any nonzero $\varrho\in\T_+(\H)$.} \smallskip

Proposition \ref{sf-1} implies the following\smallskip

\begin{property}\label{sf-2} \emph{The  function $(\rho,\sigma,\eta,\theta)\mapsto \Tr (\rho+\eta)(-\ln\sigma)-\Tr \rho(-\ln(\sigma+\theta))$
is lower semicontinuous on the set}
\begin{equation*}%\label{sf-2-set}
  \left\{(\rho,\sigma,\eta,\theta)\in[\T_+(\H)]^{\times 4}\,|\,\Tr \rho(-\ln(\sigma+\theta))<+\infty\right\}.
\end{equation*}
\end{property}

\emph{Proof.} The function $\,(\sigma,\eta)\mapsto \Tr \eta(-\ln\sigma)=D(\eta\shs\|\shs\sigma)+S(\eta)+\Tr\eta(1-\ln\Tr\eta)-\Tr\sigma\,$ is lower semicontinuous on
the set $[\T_+(\H)]^{\times 2}$ by the lower semicontinuity of the QRE and the extended von Neumann entropy. Thus, to prove the proposition it suffices to show that
the function $(\rho,\sigma,\theta)\mapsto \Tr \rho\shs(-\ln\sigma)-\Tr \rho\shs(-\ln(\sigma+\theta))$ is lower semicontinuous on
the set of all triplets $(\rho,\sigma,\theta)$ such that $\Tr \rho\shs(-\ln(\sigma+\theta))<+\infty$.

Since the finiteness of $\Tr \rho\shs(-\ln(\sigma+\theta))$ implies the finiteness of the entropy $S(\rho)$, by using representation (\ref{re-exp}) we see that
$$
\Tr\rho\shs(-\ln\sigma)-\Tr \rho\shs(-\ln(\sigma+\theta))=D(\rho\shs\|\shs\sigma)-D(\rho\shs\|\shs\sigma+\theta)+\Tr\theta.
$$
By Proposition \ref{sf-1} the function $(\rho,\sigma,\theta)\mapsto D(\rho\shs\|\shs\sigma)-D(\rho\shs\|\shs\sigma+\theta)$ is lower semicontinuous on
the set all triplets $(\rho,\sigma,\theta)$ such that $D(\rho\shs\|\shs\sigma+\theta)<+\infty$, which contains
the set of all triplets $(\rho,\sigma,\theta)$ such that $\Tr \rho\shs(-\ln(\sigma+\theta))<+\infty$. $\Box$ \smallskip

\begin{corollary}\label{sf-2}  \emph{Let $\{\rho^1_n\}$, $\{\rho^2_n\}$, $\{\sigma^1_n\}$ and $\{\sigma^2_n\}$ be sequences of  operators in $\T_+(\H)$ converging, respectively,
to operators $\rho^1_0$, $\rho^2_0$, $\sigma^1_0$ and $\sigma^2_0$ such that $\,\rho^1_n\geq \rho^2_n$ and $\,\sigma^1_n\leq \sigma^2_n$ for all $\,n\geq0$. If $\,\Tr \rho^1_0(-\ln \sigma_0^1)<+\infty\,$ then $\,\Tr \rho^2_0(-\ln \sigma_0^2)<+\infty\,$ and}\footnote{The symbol "$\djn$" is defined in (\ref{dj}).}
\begin{equation*}%\label{c-11}
\djn(\{\Tr \rho^2_n(-\ln \sigma_n^2)\})\leq\djn(\{\Tr \rho^1_n(-\ln \sigma_n^1)\}).
\end{equation*}
\emph{In particular, if
\begin{equation*}%\label{c-11+}
\lim_{n\to+\infty}\Tr \rho^1_n(-\ln \sigma_n^1)=\Tr \rho^1_0(-\ln \sigma_0^1)<+\infty
\end{equation*}
then}
\begin{equation*}%\label{c-22+}
\lim_{n\to+\infty}\Tr \rho^2_n(-\ln \sigma_n^2)=\Tr \rho^2_0(-\ln \sigma_0^2)<+\infty.
\end{equation*}
\end{corollary}

A direct proof of Corollary \ref{sf-2} obtained in \cite[the Appendix]{DTL}  requires a lot of technical efforts.

\subsubsection{The function $(\{p_i,\rho_i\},\{q_i,\sigma_i\})\mapsto \!D\left(\sum_ip_i\rho_i\|\shs\sum_iq_i\sigma_i\right)$}

By using the joint convexity of the QRE (in the form of inequality (\ref{JC++})) and identities (\ref{D-mul}) and (\ref{D-c-id}) it is easy to show that
$$
D\left.\left(\sum_ip_i\rho_i\shs\right\|\shs\sum_iq_i\sigma_i\right)\leq\sum_i p_iD(\rho_i\|\shs\sigma_i)+D_{\rm KL}(\{p_i\}\|\{q_i\})
$$
for any ensembles $\{p_i,\rho_i\}$ and $\{q_i,\sigma_i\}$ of quantum states in $\S(\H)$, where
$$
D_{\rm KL}(\{p_i\}\|\{q_i\})\doteq \sum_i p_i \ln (p_i/q_i)
$$
is the Kullback-Leibler divergence between the probability distributions $\{p_i\}$ and $\{q_i\}$ (it is assumed that $D_{\rm KL}(\{p_i\}\|\{q_i\})=+\infty$ if there is $i$ such that $p_i\neq0$ and $q_i=0$.) \cite{C&T}. Indeed, using identities (\ref{D-mul}) and (\ref{D-c-id}) it is easy to show that
\begin{equation}\label{sim-exp}
 \sum_i p_iD(\rho_i\|\shs\sigma_i)+D_{\rm KL}(\{p_i\}\|\{q_i\})=\sum_i D(p_i\rho_i\|\shs q_i\sigma_i).
\end{equation}

Theorem \ref{main}A implies the following \smallskip

\begin{property}\label{new-fun} \emph{The nonnegative function
\begin{equation*}%\label{new-fun+}
 (\{p_i,\rho_i\},\{q_i,\sigma_i\})\mapsto \sum_i p_iD(\rho_i\|\shs\sigma_i)+D_{\rm KL}(\{p_i\}\|\{q_i\})-D\!\left(\left.\sum_ip_i\rho_i\shs\right\|\shs\sum_iq_i\sigma_i\right)
\end{equation*}
is lower semicontinuous on the set
\begin{equation*}%\label{P-set++}
\left\{(\{p_i,\rho_i\},\{q_i,\sigma_i\})\in\P(\H)\times\P(\H)\left| \,D\!\left(\sum_ip_i\rho_i\shs\right\|\left.\sum_iq_i\sigma_i\right)<+\infty \right\}\right.
\end{equation*}
with respect to the $D_0$-convergence in $\P(\H)\times\P(\H)$.}\footnote{$\P(\H)$ denotes the set of all discrete ensembles of states on $\H$. The $D_0$-convergence is defined in Section 3.3.1.}
\end{property}\smallskip

\emph{Proof.} It suffices to use (\ref{sim-exp})
and to apply Proposition \ref{c-sum} in Section 3.3.2. $\Box$ \smallskip

Proposition \ref{new-fun} allows us to generalize the claims of Corollary \ref{j-c-g-c} in Section 3.3.1.
\smallskip

\begin{corollary}\label{new-fun-c} \emph{Let $\{\{p^n_i,\rho^n_i\}\}_n$ and $\{\{q^n_i,\sigma^n_i\}\}_n$ be sequences of ensembles in $\P(\H)$ $D_0$-converging to  ensembles $\{p^0_i,\rho^0_i\}$ and $\{q^0_i,\sigma^0_i\}$ such that $\sum_i p^0_iD(\rho^0_i\|\sigma^0_i)<+\infty$ and $D_{\rm KL}(\{p^0_i\}\|\{q^0_i\})<+\infty$. Then}\footnote{The symbol "$\djn$" is defined in (\ref{dj}).}
$$
\djn\left(\left\{D\!\left(\sum_ip^n_i\rho^n_i\right\|\left.\sum_iq^n_i\sigma^n_i\right)\right\}\right)\leq \djn\left(\left\{\sum_i p^n_iD(\rho^n_i\|\shs\sigma^n_i)+D_{\rm KL}(\{p^n_i\}\|\{q^n_i\})\right\}\right).
$$
\emph{In particular, if
\begin{equation}\label{new-fun-c+}
\lim_{n\to+\infty}\sum_i p^n_iD(\rho^n_i\|\sigma^n_i)=\sum_i p^0_iD(\rho^0_i\|\sigma^0_i)<+\infty
\end{equation}
and
\begin{equation}\label{new-fun-c++}
\lim_{n\to+\infty}D_{\rm KL}(\{p^n_i\}\|\{q^n_i\})=D_{\rm KL}(\{p^0_i\}\|\{q^0_i\})<+\infty
\end{equation}
then}
\begin{equation}\label{new-fun-c+++}
\lim_{n\to+\infty}D\!\left(\sum_ip^n_i\rho^n_i\shs\right\|\left.\sum_iq^n_i\sigma^n_i\right)=D\!\left(\sum_ip^0_i\rho^0_i\right\|\left.\sum_iq^0_i\sigma^0_i\right)<+\infty.
\end{equation}
\emph{If $\rho^n_i\rho^n_j=\rho^n_i\sigma^n_j=\sigma^n_i\sigma^n_j=0$ for all $i\neq j$ and any $n$ then
(\ref{new-fun-c+}) and (\ref{new-fun-c++}) are necessary conditions for (\ref{new-fun-c+++}).}
\end{corollary}\smallskip

\emph{Proof.} The main claim follows, by Lemma \ref{vsl} in Section 2.1, from Proposition \ref{new-fun} and the lower semicontinuity of the function
$(\{p_i,\rho_i\},\{q_i,\sigma_i\})\mapsto D(\sum_i p_i\rho_i\|\shs\sum_i q_i\sigma_i)$.

If $\rho^n_i\rho^n_j=\rho^n_i\sigma^n_j=\sigma^n_i\sigma^n_j=0$ for all $i\neq j$ and any $n$ then identity (\ref{D-sum}) implies that
$$
D\!\left(\sum_ip^n_i\rho^n_i\shs\right\|\left.\sum_iq^n_i\sigma^n_i\right)=\sum_i p^n_iD(\rho^n_i\|\shs\sigma^n_i)+D_{\rm KL}(\{p^n_i\}\|\{q^n_i\})\quad \forall n.
$$
So, the last claim follows, by Lemma \ref{vsl} in Section 2.1, from the lower semicontinuity of the function
$(\{p_i,\rho_i\},\{q_i,\sigma_i\})\mapsto \sum_i p_iD(\rho_i\|\shs\sigma_i)$ and the Kullback-Leibler divergence.$\Box$

\smallskip

\begin{example}\label{e-1} Let $\{\rho_i\}$ and $\{\sigma_i\}$ be countable collections of quantum states
such that $\sup_iD(\rho_i\|\sigma_i)<+\infty$. If
$\{\{p^n_i\}_i\}_n$ and $\{\{q^n_i\}_i\}_n$ are sequences of probability distributions
converging, respectively, to probability distributions  $\{p^0_i\}$ and $\{q^0_i\}$ such that condition (\ref{new-fun-c++}) holds
then Corollary \ref{new-fun-c} implies that
\begin{equation*}%\label{new-fun-c+++}
\lim_{n\to+\infty}D\!\left(\sum_ip^n_i\rho_i\shs\right\|\left.\sum_iq^n_i\sigma_i\right)=D\!\left(\sum_ip^0_i\rho_i\shs\right\|\left.\sum_iq^0_i\sigma_i\right)<+\infty.
\end{equation*}
\end{example}

\subsubsection{The function $(\mu,\nu)\mapsto \!D\left(\int \rho(x)\mu(dx)\|\shs\int \rho(x)\nu(dx)\right)$}

%Motivating by Example \ref{e-1} in Section we may conjecture the validity of its "continuous" version which can be formulated as follows.

Assume that $\rho(x)$ is a $\S(\H)$-valued measurable function on a separable metric space $X$ and
that $\{\mu_n\}$ and $\{\nu_n\}$ are sequences of probability measures on $X$ converging (in some sense) to probability measures
$\mu_0$ and $\nu_0$ such that
\begin{equation}\label{b-cond}
\lim_{n\to+\infty}D_{\rm KL}(\mu_n\|\nu_n)=D_{\rm KL}(\mu_0\|\nu_0)<+\infty,
\end{equation}
where
$$
D_{\rm KL}(\mu\|\nu)=\int_X \ln\left(\frac{d\mu}{d\nu}\right)\mu(dx)
$$
is the Kullback-Leibler divergence between probability measures $\mu$ and $\nu$,
$\frac{d\mu}{d\nu}$ denotes the  Radon-Nikodym derivative of
$\mu$ w.r.t. $\nu$ (if $\mu$ is not absolutely continuous w.r.t. $\nu$ then
$D_{\rm KL}(\mu\|\nu)=+\infty$) \cite{C&T,K-L+,K-L++}.

Motivating by Example \ref{e-1} in Section 4.1.5 (with $\sigma_i=\rho_i$) we conjecture that condition (\ref{b-cond}) implies that
\begin{equation}\label{b-conv}
\begin{array}{c}
\displaystyle\lim_{n\to+\infty} D\left(\left.\int_X \rho(x)\mu_n(dx)\right\|\shs\int_X \rho(x)\nu_n(dx)\right)\\\\\displaystyle\qquad\qquad\quad = D\left(\left.\int_X \rho(x)\mu_0(dx)\right\|\shs\int_X \rho(x)\nu_0(dx)\right)<+\infty.
\end{array}
\end{equation}

By using the main claim of Corollary \ref{new-fun-c} and a simple approximation technique one can prove the following\smallskip

\begin{property}\label{vn-fun} \emph{Let $\rho(x)$ be a $\S(\H)$-valued continuous function on a separable metric space $X$.
Let $\{\mu_n\}$ and $\{\nu_n\}$ be sequences of probability measures setwise converging\footnote{It means that $\mu_n(A)$ and $\nu_n(A)$ tends to $\mu_0(A)$ and $\nu_0(A)$ for any Borel subset $A$ of $X$ \cite{Bog}.} to probability measures
$\mu_0$ and $\,\nu_0$ such that $\mu_n$ is absolutely continuous w.r.t. $\nu_n$ for all $n\geq0$ and
the family $\left\{\frac{d\mu_n}{d\nu_n}\right\}_{n\geq 0}$ of the  Radon-Nikodym derivatives is uniformly bounded and uniformly equicontinuous on $X$, i.e.
\begin{equation}\label{t-cond}
\sup_{n\geq 0}\sup_{x\in X}\frac{d\mu_n}{d\nu_n}(x)<+\infty,\quad \sup_{n\geq 0}\sup_{d(x_1,x_2)\leq\varepsilon} \left|\frac{d\mu_n}{d\nu_n}(x_1)-\frac{d\mu_n}{d\nu_n}(x_2)\right|=o(1)\quad \textit{as}\;\; \varepsilon\to0,
\end{equation}
where $d(\cdot,\cdot)$ is the metric on $X$, then (\ref{b-cond}) implies (\ref{b-conv}). }
\end{property}\smallskip

\emph{Proof.} For given $\varepsilon>0$ take a countable decomposition  $\{X_i\}$ of $X$ into disjoint Borel subsets
with the diameter not exceeding $\varepsilon$.

For each $n$ consider the ensembles $\{p^{n}_i,\rho^{n}_i\}$ and $\{q^{n}_i,\sigma^{n}_i\}$ of states in $\S(\H)$, where
$$
p^{n}_i=\mu_n(X_i),\;\;\rho^{n}_i=\frac{1}{p^{n}_i}\int_{X_i}\!\rho(x)\mu_n(dx),\;\; q^{n}_i=\nu_n(X_i),\;\;\sigma^{n}_i=\frac{1}{q^{n}_i}\int_{X_i}\!\rho(x)\nu_n(dx).
$$
If $\mu_n(X_i)=0$ (correspondingly, $\nu_n(X_i)=0$) we assume that $p^n_i=0$ and $\rho^n_i=\tau$ (correspondingly, $q^n_i=0$ and $\sigma^n_i=\tau$), where $\tau$
is any state in $\S(\H)$. Note that
$q^n_i=0$ implies $p^n_i=0$ by the condition.

The setwise convergence of the  sequences $\{\mu_n\}$ and $\{\nu_n\}$ to the probability measures $\mu_0$ and $\nu_0$ implies the $D_0$-convergence
of the sequences $\{\{p^{n}_i,\rho^{n}_i\}_i\}_n$ and $\{\{q^{n}_i,\sigma^{n}_i\}_i\}_n$  to the ensembles  $\{p^{0}_i,\rho^{0}_i\}$ and $\{q^{0}_i,\sigma^{0}_i\}$.\footnote{The $D_0$-convergence is defined in Section 3.3.1.}

To simplify notation denote the function $\frac{d\mu_n}{d\nu_n}$ on $X$ by $f_n$.
Condition (\ref{t-cond}) implies that
\begin{equation}\label{t-cond+}
\sup_{n\geq 0}\sup_{x\in X}f_n(x)\leq b\quad \textrm{and} \quad\sup_{n\geq 0}\sup_{d(x_1,x_2)\leq\varepsilon}\left|f_n(x_1)-f_n(x_2)\right|\leq \alpha(\varepsilon),
\end{equation}
where $b\in\mathbb{R}_+$ and $\alpha(\varepsilon)$ is a function tending to zero as $\,\varepsilon\to0^+$.

Let $\omega_{\eta,b}(\varepsilon)\doteq \sup\{|\eta(x_1)-\eta(x_2)\,|\,x_1,x_2\in[0,b],\,|x_1-x_2|\leq\varepsilon\}$
be the modulus of continuity of the function $\eta(x)=-x\ln x$ on $[0,b]$ (it is assumed that $\eta(0)=0$).

We will show that
\begin{equation}\label{A-r}
\sum_i p^n_iD(\rho^n_i\|\sigma^n_i)\leq \omega_{\eta,b}(\alpha(\varepsilon))\quad \forall n\geq 0
\end{equation}
and
\begin{equation}\label{B-r}
0\leq D_{\rm KL}(\mu_n\|\nu_n)-D_{\rm KL}(\{p^n_i\}\|\{q^n_i\})\leq \omega_{\eta,b}(\alpha(\varepsilon))\quad \forall n\geq 0.
\end{equation}

For any given $n\geq0$ let $a^n_i=\inf_{x\in X_i}f_n(x)$ and $b^n_i=\sup_{x\in X_i}f_n(x)$. Since for each $i$ the diameter of the set $X_i$
does not exceed $\varepsilon$, it follows from (\ref{t-cond+}) that
\begin{equation}\label{t-cond++}
b^n_i-a^n_i\leq \alpha(\varepsilon).
\end{equation}

Since $p_i=\mu_n(X_i)=\int_{X_i}f_n(x)\nu_n(dx)$ and $q_i=\nu_n(X_i)$, we have
\begin{equation}\label{p-q}
a^n_i\leq p^n_i/q^n_i\leq b^n_i\quad \forall i:q^n_i\neq0.
\end{equation}

It follows from (\ref{t-cond++}) and (\ref{p-q}) that
\begin{equation}\label{p-q+}
\sup_{x\in X_i}|f_n(x)- p^n_i/q^n_i|\leq \alpha(\varepsilon)\quad  \forall i:q^n_i\neq0.
\end{equation}

To prove (\ref{A-r}) note that $\nu_n/q^n_i$ is a probability measure on $X_i$ for each $i$ and $n$ such that $q_i^n\neq0$. So, by using Lemma \ref{JC-int} below and identities
(\ref{D-mul}) and (\ref{D-c-id}) we obtain
$$
\begin{array}{rl}
\displaystyle p^n_iD(\rho^n_i\|\sigma^n_i)\!\!&\displaystyle=\;p^n_iD\!\left(\left.\int_{X_i} \displaystyle \frac{q^n_i}{p^n_i}\rho(x)f_n(x)\frac{\nu_n(dx)}{q^n_i}\,\right\|\shs\int_{X_i}\rho(x)\frac{\nu_n(dx)}{q^n_i}\right)\\\\
&\displaystyle\leq \;\int_{X_i} D(q^n_i\rho(x)f_n(x)\|\shs p^n_i\rho(x))\frac{\nu_n(dx)}{q^n_i}=\int_{X_i} f_n(x)\ln \left(\frac{q^n_if_n(x)}{p_i^n}\right)\nu_n(dx)
\\\\
&\displaystyle=\;\int_{X_i}\left(\eta\left(\frac{p^n_i}{q_i^n}\right)-\eta(f_n(x))\right)\nu_n(dx)\leq q^n_i\omega_{\eta,b}(\alpha(\varepsilon))
\end{array}
$$
for each $i$ and $n$ such that $p_i^n\neq0$, where the last inequality follows from (\ref{p-q+}).
This inequality implies (\ref{A-r}).

The left inequality in (\ref{B-r}) follows from by the monotonicity of the Kullback-Leibler divergence under the positive linear map
$\mu\mapsto \{\mu(X_i)\}_i$ from the space of all signed Borel measures on $X$ to the space $\ell_1$
determined by the decomposition $\{X_i\}_i$ of $X$.

To prove the right inequality in (\ref{B-r}) note that (\ref{p-q+}) implies
$$
\begin{array}{c}
\displaystyle D_{\rm KL}(\mu_n\|\nu_n)=-\int_{X}\eta(f_n(x))\nu_n(dx)\leq\sum_i\int_{X_i}\left(-\eta\left(\frac{p_i^n}{q_i^n}\right)+\omega_{\eta,b}(\alpha(\varepsilon))\right)\nu_n(dx)
\\\\=D_{\rm KL}(\{p^n_i\}\|\{q^n_i\})+\omega_{\eta,b}(\alpha(\varepsilon)).
\end{array}
$$

Inequalities (\ref{A-r}) and (\ref{B-r}) show that
$$
\sup_{n\geq0} \left|\sum_i p^n_iD(\rho^n_i\|\sigma^n_i)+D_{\rm KL}(\{p^n_i\}\|\{q^n_i\})-D_{\rm KL}(\mu_n\|\nu_n)\right|\leq \omega_{\eta,b}(\alpha(\varepsilon)).
$$
So, it follows from condition (\ref{b-cond})  that
$$
\djn\left(\left\{\sum_i p^n_iD(\rho^n_i\|\sigma^n_i)+D_{\rm KL}(\{p^n_i\}\|\{q^n_i\})\right\}_n\right)\leq 2\omega_{\eta,b}(\alpha(\varepsilon)).
$$
Hence, by Corollary \ref{new-fun-c} in Section 4.1.5 we have
$$
\djn\left(\left\{D\!\left(\sum_ip^n_i\rho^n_i\right\|\left.\sum_iq^n_i\sigma^n_i\right)\right\}_n\right)\leq 2\omega_{\eta,b}(\alpha(\varepsilon)).
$$
Since $\omega_{\eta,b}(\alpha(\varepsilon))$ can be made arbitrarily small by choosing sufficiently small $\varepsilon$,  we obtain
(\ref{b-conv}) by noting that
\begin{equation*}%\label{av-s}
  \sum_ip_i^n\rho_i^n=\int_X \rho(x)\mu_n(dx)\quad\textrm{and}\quad\sum_iq_i^n\sigma_i^n=\int_X \rho(x)\nu_n(dx).\;\;\Box
\end{equation*}

\begin{remark}\label{vn-fun-r} The setwise convergence of the  sequences $\{\mu_n\}$ and $\{\nu_n\}$ to the measures $\mu_0$ and $\nu_0$
in Proposition \ref{vn-fun} can be relaxed to the weak convergence provided that for any $\varepsilon>0$ there exists a countable decomposition $\{X_i\}$ of the  space $X$ into disjoint Borel subsets  with the diameter $\leq \varepsilon$ such that $\mu_0(\partial X_i)=\nu_0(\partial X_i)=0$ for all $i$, where $\partial X_i$ is the boundary of $X_i$.
This follows from the above proof and the Portmanteau theorem \cite{Bill,Bog}.

The condition (\ref{t-cond}) is quite restrictive, but it seems that it can be relaxed
by using more subtle estimates in the proof of inequalities (\ref{A-r}) and (\ref{B-r}).
\end{remark}\smallskip

\begin{example}\label{e-2}
Let $\H$ be the Hilbert space describing $n$-mode quantum oscillator and $\S_\mathrm{cl}(\H)$ is the set of classical states -- the convex closure
of the family $\{|\bar{z}\rangle\langle \bar{z}| \}_{\bar{z}\in\mathbb{C}^n}$ of coherent states \cite{H-SCI,IQO}. Each state
$\rho$ in $\S_\mathrm{cl}(\H)$ can be represented as
\begin{equation}\label{P-rep}
 \rho=\int_{\mathbb{C}^n}|\bar{z}\rangle\langle \bar{z}| \mu_{\rho}(dz_1...dz_n),\quad  \bar{z}=(z_1,...,z_n),
\end{equation}
where $\mu_{\rho}$ is a Borel probability measure on $\mathbb{C}^n$ that can be called \emph{representing measure} for $\rho$.  Representation (\ref{P-rep}) is the Glauber-Sudarshan $P$-representation, since
generally it is written as
\begin{equation*}%\label{P-rep+}
 \rho=\int_{\mathbb{C}^n}|\bar{z}\rangle\langle \bar{z}| P_{\rho}(\bar{z})dz_1...dz_n,
\end{equation*}
where $P_{\rho}$ is the $P$-function of a state $\rho$, which in this case is nonnegative and can be treated as
a generalized probability density function on $\mathbb{C}^n$ (in contrast to the standard probability density,
the function $P_{\rho}$ may be singular, since the measure $\mu_{\rho}$ may be not absolutely continuous w.r.t. the Lebesgue measure on
$\mathbb{C}^n$) \cite{Gla,Sud}.

Proposition \ref{vn-fun} with $X=\mathbb{C}^n$ and $\rho(\bar{z})=|\bar{z}\rangle\langle \bar{z}|$
and Remark \ref{vn-fun-r} allows us to obtain  a sufficient condition for convergence of the QRE between classical states
of the $n$-mode quantum oscillator due the following property: \emph{for any sequence $\{\rho_n\}$ of classical states
converging to a classical state $\rho_0$ there exist a subsequence $\{\rho_{n_k}\}$ and a sequence $\{\mu_k\}$
of Borel probability measures on $\mathbb{C}^n$ weakly
converging to a Borel probability measure $\mu_0$ such that $\mu_k$ is a representing measure for $\rho_{n_k}$
for all $\,k$ and $\mu_0$ is a representing measure for $\rho_{0}$}. This property follows
from Proposition 2 in \cite{H-Sh-2}. \smallskip
\end{example}

The following lemma contains a "continuous" version of the QRE joint convexity.\footnote{I am sure that the claim of Lemma \ref{JC-int} can be found in the literature.     I would be grateful for the corresponding reference.}\smallskip

\begin{lemma}\label{JC-int} \emph{Let $\rho(x)$ and $\sigma(x)$ be $\,\S(\H)$-valued continuous functions
on a separable metric space $X$. Then
\begin{equation}\label{JC-int+}
D\left(\left.\int_X \rho(x)\mu(dx)\shs\right\|\shs\int_X \sigma(x)\mu(dx)\right)\leq \int_X D(\rho(x)\|\sigma(x))\mu(dx)
\end{equation}
for any Borel probability measure $\mu$ on $X$.}
\end{lemma}\smallskip

\textbf{Note:} Since the QRE is lower semicontinuous, the function $x\mapsto D(\rho(x)\|\sigma(x))$ is lower semicontinuous on $X$ and, hence, measurable w.r.t. the Borel
$\sigma$-algebra on $X$. So, the r.h.s. of (\ref{JC-int+}) is well defined.\smallskip

\emph{Proof.}  The joint convexity of the QRE implies that (\ref{JC-int+}) holds for any purely atomic (discrete) probability
measure $\mu$, i.e. a measure of the form $\sum_ip_i\delta(x_i)$, where $\{p_i\}$ is a probability distribution,
$\{x_i\}$ is a finite or countable subset of $X$ and $\delta(x)$ denotes the Dirac measure concentrating at $x$.

For a given probability measure $\mu$ on $X$ it is easy\footnote{This can be done by taking for each $n$ a countable decomposition $\{X_i^n\}$ of the separable space $X$ into
disjoint subsets  with the diameter $\leq1/n$, choosing a proper point $x_i$ in each subset and using the Portmanteau theorem
to prove that the sequence of measures $\mu_n=\sum_i \mu(X_i)\delta(x_i)$ weakly converges to the measure $\mu$ \cite{Bill,Bog}.} to construct a sequence $\{\mu_n\}$ of purely atomic probability measures on $X$
weakly converging to a given probability measure $\mu$ on $X$ such that
\begin{equation}\label{p-in}
\int_X D(\rho(x)\|\sigma(x))\mu_n(dx)\leq \int_X D(\rho(x)\|\sigma(x))\mu(dx)\quad \forall n.
\end{equation}
By using the definition of the weak convergence (cf. \cite{Bill,Bog}) and Lemma \ref{D-A} in Section 2.1 it is easy to prove
that
$$
\lim_{n\to+\infty}\int_X \rho(x)\mu_n(dx)=\int_X \rho(x)\mu(dx)\quad \textrm{and}\quad \lim_{n\to+\infty}\int_X \sigma(x)\mu_n(dx)=\int_X \sigma(x)\mu(dx)
$$
So, the lower semicontinuity of the QRE shows that
$$
\liminf_{n\to+\infty}D\left(\left.\int_X \rho(x)\mu_n(dx)\right\|\shs\int_X \sigma(x)\mu_n(dx)\right)\geq D\left(\left.\int_X \rho(x)\mu(dx)\right\|\shs\int_X \sigma(x)\mu(dx)\right).
$$
Since (\ref{JC-int+}) holds with $\mu=\mu_n$, the last inequality and (\ref{p-in}) implies (\ref{JC-int+}). $\Box$

\subsubsection{Quantum mutual information} The lower semicontinuity of the function
$(\rho,\sigma)\mapsto D(\rho\shs\|\shs\sigma)-D(\Phi(\rho)\shs\|\shs\Phi(\sigma))$
proved in this article in
the general form was established before in the special case when
$\rho$ is a state of a composite quantum system $A_1...A_n$,
$\sigma=\rho_{A_1}\otimes...\otimes\rho_{A_n}$ and $\Phi$ is a local channel \cite{LSE}.
In this case  $D(\rho\shs\|\shs\sigma)=I(A_1\!:...:\!A_n)_{\rho}$ is the \emph{quantum mutual
information} (QMI) of a state $\rho$, which describes the total correlation of this state \cite{L-MI,NQD}.

The lower semicontinuity of the QMI decrease under action of a local channel is related (due to the Stinespring representation)
to the lower semicontinuity of the quantum conditional mutual
information (\ref{QCMI}) proved in \cite[Section 6]{CMI}. In Section 5 in \cite{LSE} this property is applied to obtain
new convergence conditions for the QMI (conditional and unconditional) and for the squashed entanglement, whose effectiveness
is shown by concrete examples. All these results can be treated as applications of Theorem \ref{main} in Section 3.1 as well.

Here we present one observation that was not mentioned in \cite{LSE}.\smallskip\pagebreak

\begin{property}\label{QMI-appl} \emph{Let $A_1...A_n$ be a $n$-partite quantum system, $n>2$, and\break $B_1=A_{i^1_1}...A_{i^1_{k(1)}}$,.., $B_m=A_{i^m_1}...A_{i^m_{k(m)}}$
its subsystems determined by disjoint subsets $\,\{i^1_1,..,i^1_{k(1)}\}$,.., $\{i^m_1,..,i^m_{k(m)}\}$  of $\,[1,n]\cap\mathbb{N}$, $m<n$}.\footnote{We do not assume that
$B_1...B_m=A_1...A_n$ in general.}

\emph{The nonnegative function $\,\rho\mapsto I(A_1\!:...:\!A_n)_{\rho}-I(B_1\!:...:\!B_m)_{\rho}\,$ is lower semicontinuous
on the set}
\begin{equation}\label{setB}
\left\{\rho\in\S(\H_{A_1..A_n})\,|\,I(B_1\!:...:\!B_m)_{\rho}<+\infty\right\}.
\end{equation}

\emph{If the function $\rho\mapsto I(A_1\!:...:\!A_n)_{\rho}$ is continuous
on some subset $\S_0$ of $\S(\H_{A_1..A_n})$ then the function $\rho\mapsto I(B_1\!:...:\!B_m)_{\rho}$
is continuous on $\S_0$.}
\end{property}\smallskip

\emph{Proof.} We may assume w.l.o.g. that $\bigcup_{j=1}^m\{i^j_1,..,i^j_{k(j)}\}=[1,n']\cap\mathbb{N}$ for some $n'\leq n$.

Theorem \ref{main}A implies that the function
$\,\rho\mapsto I(A_1\!:...:\!A_n)_{\rho}-I(A_1\!:...:\!A_{n'})_{\rho}\,$ is lower semicontinuous
on the set
$$
\mathfrak{A}=\left\{\rho\in\S(\H_{A_1..A_n})\,|\,I(A_1\!:...:\!A_{n'})_{\rho}<+\infty\right\}.
$$
Since
\begin{equation}\label{I-iden}
I(A_1\!:...:\!A_{n'})_{\rho}=I(B_1\!:...:\!B_m)_{\rho}+\sum_{j=1}^m I(A_{i^j_1}\!:\!...\!:\!A_{i^j_{k(j)}})_{\rho},
\end{equation}
the lower semicontinuity of the QMI implies that the function $\,\rho\mapsto I(A_1\!:...:\!A_{n'})_{\rho}-I(B_1\!:...:\!B_m)_{\rho}\,$ is lower semicontinuous
on the set in (\ref{setB}), which we denote by $\mathfrak{B}$.

Thus, the function $f(\rho)=I(A_1\!:...:\!A_n)_{\rho}-I(B_1\!:...:\!B_m)_{\rho}$ is  lower semicontinuous  on the set $\mathfrak{A}\subseteq\mathfrak{B}$.

It is clear that $f(\rho)=+\infty$ for any $\rho$ in $\mathfrak{B}\setminus\mathfrak{A}$. So, to prove
the lower semicontinuity of $f$ on $\mathfrak{B}$ it suffices to show that
$\lim_{n\to+\infty}f(\rho_n)=+\infty$
for any sequence $\{\rho_n\}\subset\mathfrak{A}$ converging to a state $\rho_0\in\mathfrak{B}\setminus\mathfrak{A}$. This can be done
by noting that identity (\ref{I-iden}) implies that
$$
f(\rho_n)\geq \sum_{j=1}^m I(A_{i^j_1}\!:\!...\!:\!A_{i^j_{k(j)}})_{\rho_n}\quad \forall n,
$$
since the r.h.s. of this inequality tends to $\,\sum_{j=1}^m I(A_{i^j_1}\!:\!...\!:\!A_{i^j_{k(j)}})_{\rho_0}=+\infty\,$ by the lower semicontinuity of
QMI. \

The second claim of the proposition follows from the first one by Lemma \ref{vsl} in Section 2.1. $\Box$
\smallskip

By Proposition \ref{QMI-appl} the local continuity (convergence) of the total correlation in the
"large" composite system $A_1..A_n$ implies the local continuity (convergence) of the total correlation in
any composite system $B_1...B_m$  whose components $B_1$,..,$B_m$ are obtained by combining some of the subsystems $A_1$,..,$A_n$.
For example, the local continuity of $I(A\!:\!B\!:\!C)$ implies the local continuity of
$I(A\!:\!B)$, $I(B\!:\!C)$, $I(A\!:\!C)$, $I(A\!:\!BC)$, $I(AB\!:\!C)$ and $I(AC\!:\!B)$.\smallskip

\begin{remark}\label{CMI-r} The claim of Proposition \ref{QMI-appl} is also valid for the multipartite quantum conditional
mutual information, i.e. it holds with the functions $I(A_1\!:...:\!A_n)_{\rho}$ and $I(B_1\!:...:\!B_m)_{\rho}$ replaced by
$I(A_1\!:...:\!A_n|\shs C)_{\rho}$ and $I(B_1\!:...:\!B_m|\shs C)_{\rho}$. This can be shown by the similar arguments using Theorem 3 in \cite{LSE} and the lower semicontinuity of the multipartite quantum conditional
mutual information \cite{CMI}.
\end{remark}

\subsubsection{Information gain of local measurements}

The information gain $\mathrm{IG}(\M,\rho)$ of a quantum measurement described by a Positive Operator Values Measure (POVM) $\M=\{M_i\}$ on $\H_A$ at a state $\rho$ in $\S(\H_A)$ can be defined as
the mutual information $I(\Psi_{\M},\rho)$ of the channel
$\Psi_{\M}(\rho)=\sum_{i}[\Tr M_i\rho]|i\rangle\langle i|$
from $\T(\H_A)$ to $\T(\H_{E})$ at the state $\rho$, where $\{|i\rangle\}$ is a basic in a Hilbert space $\H_E$ such that $\dim\H_E=\mathrm{card}\{M_i\}$ \cite{GIB,IGSI+}, i.e.
$$
\mathrm{IG}(\M,\rho)=I(E\!:\!R)_{\Psi_{\M}\otimes \id_{R}(\bar{\rho})},
$$
where $\bar{\rho}$ is a purification in $\S(\H_{AR})$ of the state $\rho$. \smallskip

Assume now that $\rho$ is a state of a bipartite system $AB$ and $\M=\{M_i\}$ is a POVM on $\H_A$. Then $\M\otimes I_B=\{M_i\otimes I_B\}$ is a  POVM on $\H_{AB}$ describing
a local measurement  applied to the part $A$ of a state of the system $AB$. It is easy to see that
$$
\mathrm{IG}(\M\otimes I_B,\rho)\leq \mathrm{IG}(\M,\rho_A).
$$
Indeed, since $\Psi_{\M\otimes I_B}=\Psi_{\M}\circ\Theta$, where $\Theta=\Tr_B(\cdot)$, the second chain rule for the
mutual information of a channel (cf.\cite{H-SCI,Wilde}) implies that
\begin{equation}\label{IG-tmp}
\mathrm{IG}(\M,\rho_A)-\mathrm{IG}(\M\otimes I_B,\rho)=I(\Psi_{\M},\Theta(\rho))-I(\Psi_{\M}\circ\Theta,\rho)\geq 0.
\end{equation}

Theorem \ref{main}A in Section 3.1 implies the following \smallskip

\begin{property}\label{IG-LM} \emph{Let $\,\M=\{M_i\}$ be a POVM on $\H_A$. Then the nonnegative function
$$
\rho\mapsto\mathrm{IG}(\M,\rho_A)-\mathrm{IG}(\M\otimes I_B,\rho)
$$
is lower semicontinuous on the set $\{\rho\in\S(\H_{AB})\,|\,\mathrm{IG}(\M\otimes I_B,\rho)<+\infty\}$}.
\end{property} \smallskip

\emph{Proof.} Basic results of the purification theory (cf. \cite{H-SCI,Wilde}) imply that for any sequence $\{\rho_n\}$ of  states in $\S(\H_{AB})$
converging to a state $\rho_0$ there is a sequence $\{\bar{\rho}_n\}$ of pure states in $\S(\H_{ABR})$
converging to a pure state $\bar{\rho}_0$ such that $\Tr_R\bar{\rho}_n=\rho_n$ for all $n\geq0$.

Using this observation, expression (\ref{chain-2}) and the equality in (\ref{IG-tmp}) it is easy to obtain
the claim of the proposition from Theorem \ref{main}A. $\Box$ \smallskip\pagebreak

\begin{corollary}\label{IG-LM-c} \emph{Let $\M=\{M_i\}$ be a POVM on $\H_A$. If $\,\{\rho_n\}$ is a sequence in $\S(\H_{AB})$
converging to a state $\rho_0$ such that
\begin{equation}\label{IG-lr}
\lim_{n\to+\infty}\mathrm{IG}(\M,[\rho_n]_A)=\mathrm{IG}(\M,[\rho_0]_A)
\end{equation}
then}
\begin{equation}\label{IG-lr+}
\lim_{n\to+\infty}\mathrm{IG}(\M\otimes I_B,\rho_n)=\mathrm{IG}(\M\otimes I_B,\rho_0).
\end{equation}

\emph{Limit relation (\ref{IG-lr}) holds provided that }
\begin{equation*}%\label{IG-lr++}
\lim_{n\to+\infty}S([\rho_n]_A)=S([\rho_0]_A)<+\infty.
\end{equation*}
\end{corollary}

\emph{Proof.} The first claim follows from Proposition \ref{IG-LM} and Lemma \ref{vsl} in Section 2.1 due to the lower semicontinuity of the function
$\rho\mapsto\mathrm{IG}(\M\otimes I_B,\rho)$. The second claim follows from Proposition 6A in \cite{LSE}. $\Box$ \smallskip

The convergence conditions for the information gain of quantum measurements in Proposition 6A in \cite{LSE}
states that (\ref{IG-lr+}) holds provided that either $\,\mathrm{card}\shs\M<+\infty\,$ or
\begin{equation*}%\label{IG-lr++}
\lim_{n\to+\infty}S(\rho_n)=S(\rho_0)<+\infty.
\end{equation*}

Corollary \ref{IG-LM-c} gives additional sufficient conditions for (\ref{IG-lr+}) obtained by taking the local structure of the POVM $\,\M\otimes I_B$ into account.

\subsubsection{Quantum conditional relative entropy}

The quantum conditional relative entropy between states $\rho$ and $\sigma$ of a composite system $AB$ is defined by Capel, Lucia and Perez-Garcia as
$$
D_A(\rho\shs\|\shs\sigma)=D(\rho\shs\|\shs\sigma)-D(\rho_B\shs\|\shs\sigma_B).
$$
In article \cite{CondRE}, where this notion was introduced, one can find a number of its interesting
properties. The motonicity of the QRE shows that $D_A(\rho\shs\|\shs\sigma)\geq0$ for any $\rho$ and $\sigma$.\smallskip

Theorem \ref{main} and Corollary \ref{NMP-c} in Section 3 imply directly the following\smallskip

\begin{property}\label{QCRE-appl} \emph{The nonnegative function $(\rho,\sigma)\mapsto D_A(\rho\shs\|\shs\sigma)$ is lower semicontinuous
on the set}
$$
\left\{\left.(\rho,\sigma)\in [\S(\H_{AB})]^{\times2}\,\right|\,D(\rho_B\shs\|\shs\sigma_B)<+\infty\right\}.
$$

\emph{If the function $\,(\rho,\sigma)\mapsto D(\rho\shs\|\shs\sigma)\,$ is continuous
on some subset $\S_0$ of $\,[\S(\H_{AB})]^{\times2}$ then the function $\,(\rho,\sigma)\mapsto D_A(\rho\shs\|\shs\sigma)\,$
is continuous on $\S_0$.}
\end{property}\smallskip

\subsection{On properties of channels mapping the Gibbs state of the input system into the Gibbs state of the output system}

\subsubsection{Energy-limited property}

Let $H_X$ be a positive operator on a Hilbert space $\H_X$ satisfying the Gibbs condition
\begin{equation}\label{G-cond}
  \Tr e^{-\beta H_X}<+\infty\quad \forall\beta>0,\quad X=A,B.
\end{equation}

We will treat $H_X$ as a Hamiltonian (energy observable) of a quantum system $X$.
So, for any state $\rho\in\S(\H_X)$ the value of $\,E_{H_{\!X}}(\rho)\doteq\Tr H_X\rho\,$ defined in (\ref{H-fun}) is the \emph{mean energy} of this  state.

Let
\begin{equation}\label{Gibbs}
\gamma_{H_{\hsh X\hsh},\beta}=e^{-\beta H_X}/\Tr e^{-\beta H_X}
\end{equation}
be the Gibbs state of a system $X$ at inverse temperature $\beta>0$ \cite{O&P,W,H-SCI}. \smallskip

We will use the following\smallskip

\begin{property}\label{dc-l} \emph{Let $H_X$ be a positive operator on $\H_X$ satisfying condition
(\ref{G-cond}), $X=A,B$.}\smallskip

A) \emph{If $\Phi$ is a quantum channel from $A$ to $B$ such that $\Phi(\gamma_{H_{\hsh A\hsh},\beta})=\gamma_{H_{\hsh B\hsh},\beta'}$ for some $\beta,\beta'>0$ then}
\begin{equation*}%\label{H-b}
\sup\left\{E_{H_{\hsh B}}(\Phi(\rho))\,|\,\rho\in\S(\H_A),\,E_{H_{\hsh A}}(\rho)\leq E\right\}<+\infty\quad \forall E>0.
\end{equation*}

B) \emph{If $\{\Phi_n\}$ is a sequence of channels from $A$ to $B$ strongly converging to a channel  $\Phi_0$
such that $\,\Phi_n(\gamma_{H_{\hsh A\hsh},\beta})=\gamma_{H_{\hsh B\hsh},\beta'_n}\,$
for some positive $\beta$, $\beta'_n$, $n\geq0$, then}
\begin{equation*}%\label{H-b-n}
\sup_{n\geq0}\shs\sup \left\{E_{H_{\hsh B}}(\Phi_n(\rho))\,|\,\rho\in\S(\H_A),\,E_{H_{\hsh A}}(\rho)\leq E\right\}<+\infty\quad \forall E>0.
\end{equation*}
\end{property}

\emph{Proof.} It suffices to prove claim B.\smallskip

Since $\Phi_n(\gamma_{H_{\hsh A\hsh},\beta})=\gamma_{H_{\hsh B\hsh},\beta'_n}$ tends to $\Phi_0(\gamma_{H_{\hsh A\hsh},\beta})=\gamma_{H_{\hsh B\hsh},\beta'_0}$ by the condition, we conclude that
$\beta'_n$ tends to $\beta'_0$. Hence $\beta'_*\doteq\min_{n\geq0}\beta'_n>0$.   \smallskip

Denote by $\C_{H_{\hsh A\hsh},E}$ the set of all states $\rho$ in $\S(\H_A)$ such that $E_{H_{\hsh A\hsh}}(\rho)\leq E$. By the monotonicity of the QRE
for any state $\rho$ in $\C_{H_{\hsh A\hsh},E}$ we have
$$
D(\Phi_n(\rho)\|\shs\gamma_{H_{\hsh B\hsh},\beta'_n})\leq D(\rho\shs\|\shs\gamma_{H_{\hsh A\hsh},\beta})\leq\beta\Tr H_A\rho+\ln\Tr e^{-\beta H_A}\leq E+\ln\Tr e^{-\beta H_A}\quad \forall n\geq 0.
$$
So, since $\gamma_{H_{\hsh B\hsh},\beta'_n}\leq C_n\gamma_{H_{\hsh B\hsh},\beta'_*}$, where $C_n=\Tr e^{-\beta'_* H}/\Tr e^{-\beta'_n H}$, by using inequality (\ref{re-ineq}) and identity (\ref{D-c-id})  we obtain
$$
D(\Phi_n(\rho)\|\shs\gamma_{H\!,\beta_*})\leq D(\Phi_n(\rho)\|\shs\gamma_{H_{\hsh B\hsh},\beta'_n})+\ln C_n\leq E+\ln\Tr e^{-\beta H_A}+\ln C_n<+\infty\quad \forall n\geq 0.
$$
Thus, $\,\sup_{n\geq0}\sup_{\rho\in\C_{H_{\hsh A\hsh},E}}D(\Phi_n(\rho)\|\shs\gamma_{H_{\hsh B\hsh},\beta_*})<+\infty\,$ because $C_n$ tends to $C_0<+\infty$ as $n\to+\infty$. Since $\Tr[\gamma_{H_{\hsh B\hsh},\beta_*}]^\lambda<+\infty$ for any $\lambda>0$,
by Proposition 3 in \cite{EC} this implies $\,\sup_{n\geq0}\sup_{\rho\in\C_{H_{\hsh A\hsh},E}}S(\Phi_n(\rho))<+\infty\,$ and hence $\,\sup_{n\geq0}\sup_{\rho\in\C_{H_{\hsh A\hsh},E}}E_{H_{\hsh B}}(\Phi_n(\rho))<+\infty$. $\Box$ \smallskip

Claim A of Proposition \ref{dc-l} states that any quantum channel $\Phi$ from $A$ to $B$
such that $\Phi(_{H_{\hsh A\hsh},\beta})=\gamma_{H_{\hsh B\hsh},\beta'}$  for some $\beta,\beta'>0$ is \emph{energy-limited}
in terms of \cite{W-EBN}, i.e. it has a finite energy amplification factor.

\subsubsection{On continuity if the mean energy and the free energy}

The \emph{free energy} of a state $\rho$ of a quantum system $X$ with finite mean energy $E_{H_{\hsh X}}(\rho)$ at inverse temperature $\beta>0$ is defined as
$$
F_{H_{\hsh  X\hsh},\beta}(\rho)=E_{H_{\hsh X}}(\rho)-\beta^{-1}S(\rho)
$$
(the finiteness of $E_{H_{\hsh X}}(\rho)$ implies the finiteness of $S(\rho)$) \cite{W}.
It is easy to see that
\begin{equation*}%\label{F-exp}
 D(\rho\shs\|\shs\gamma_{H_{\hsh  X\hsh},\beta})=\beta E_{H_{\hsh  X}}(\rho)-S(\rho)+\ln\Tr e^{-\beta H_X}=\beta(F_{H_{\hsh  X\hsh},\beta}(\rho)-F_{H_{\hsh  X\hsh},\beta}(\gamma_{H_{\hsh  X\hsh},\beta})),
\end{equation*}
where  $\gamma_{H_{\hsh  X\hsh},\beta}$ is  the Gibbs state defined in (\ref{Gibbs}).
Thus, we may define the free energy for any state $\rho$ in $\S(\H_X)$ by the expression
\begin{equation}\label{F-exp+}
 F_{H_{\hsh  X\hsh},\beta}(\rho)=\beta^{-1}D(\rho\shs\|\shs\gamma_{H_{\hsh  X\hsh},\beta})+F_{H_{\hsh  X\hsh},\beta}(\gamma_{H_{\hsh  X\hsh},\beta})=\beta^{-1}\!\left(D(\rho\shs\|\shs\gamma_{H_{\hsh  X\hsh},\beta})-\ln\Tr e^{-\beta H_X}\right).
\end{equation}
Since $\Tr[\gamma_{H_X,\beta}]^{\lambda}<+\infty$ for any $\lambda>0$,  claim 2 of Proposition 3 in \cite{EC} implies that
\begin{equation}\label{F-E-r}
\{F_{H_{\hsh  X\hsh},\beta}(\rho)<+\infty\}\quad\Leftrightarrow\quad \{E_{H_{\hsh  X}}(\rho)<+\infty\}.
\end{equation}

Both functions $E_{H_{\hsh  X\hsh}}(\rho)$ and $F_{H_{\hsh  X\hsh}\!,\beta}(\rho)$ are lower semicontinuous on $\S(\H_X)$ (the lower semicontinuity of
the free energy follows from expression  (\ref{F-exp+}) by the lower semicontinuity of the QRE).

It is known that the mean energy $E_{H_{\hsh  X\hsh}}(\rho)$  has more singular properties (as a function of $\rho$) than the
von Neumann entropy $S(\rho)$. Indeed, the function $S(\rho)$ is uniformly continuous on the set of states
$\rho$ such that $E_{H_{\hsh  X\hsh}}(\rho)\leq E$ for any $E>0$ \cite{W},\cite[Proposition 1]{EC}, but it is easy to see that the function $E_{H_{\hsh  X\hsh}}(\rho)$ is not continuous on this set
if $E$ is greater than the minimal eigenvalue of $H_X$.

At the same time, by using claim 3 of Proposition 3 in \cite{EC} it is easy to show  that
\begin{equation*}%\label{F-E-r+}
\left\{\lim_{n\to+\infty}D(\rho_n\|\shs\gamma_{H_{\hsh  X\hsh},\beta})=D(\rho_0\|\shs\gamma_{H_{\hsh  X\hsh},\beta})<+\infty\right\}\;\Leftrightarrow\; \left\{\lim_{n\to+\infty}E_{H_{\hsh  X}}(\rho_n)=E_{H_{\hsh  X}}(\rho_0)<+\infty\right\},
\end{equation*}
where  $\{\rho_n\}\subset\S(\H_X)$ is a sequence converging to a state $\rho_0$.
This and expression  (\ref{F-exp+}) show that \emph{the local continuity
of the free energy $F_{H_{\hsh  X\hsh}\!,\beta}(\rho)$ is equivalent to the local continuity
of the mean energy  $E_{H_{\hsh  X}}(\rho)$}. So, in what follows we will restrict attention to the function $E_{H_{\hsh  X}}(\rho)$.

The above equivalence relation shows that the criterion of local continuity
of the function $\rho \mapsto D(\rho\|\shs\gamma_{H_{\hsh  X\hsh},\beta})$
presented in Proposition 2 in \cite{REC} gives a criterion of local continuity
of the mean energy $E_{H_{\hsh  X}}(\rho)$. The last criterion can be proved directly by using
simple estimates and Dini's lemma.

The results of Section 3.2 allows us to obtain conditions under which the local continuity of the
function $E_{H_{\hsh  A}}$ is preserved by quantum channels from $A$ to $B$.\smallskip\pagebreak

\begin{property}\label{H-case} \emph{Let $H_X$ be a positive operator on $\H_X$ satisfying condition
(\ref{G-cond}), $X=A,B$. Let $\{\Phi_n\}$ be a sequence of channels from $A$ to $B$ strongly converging to a channel $\Phi_0$
such that
\begin{equation}\label{Phi-cond}
  \Phi_n(\gamma_{H_{\hsh  A\hsh},\beta})=\gamma_{H_{\hsh  B\hsh},\beta'_n}
\end{equation}
for some positive $\beta$, $\beta'_n$, $n\geq0$. Then
\begin{equation}\label{H-dj}
\djn(\{E_{H_{\hsh  B}}(\Phi_n(\rho_n))\} \leq (\beta/\beta'_0)\shs\djn(\{E_{H_{\hsh  A\hsh}}(\rho_n)\})
\end{equation}
for any sequence  $\{\rho_n\}$ converging to a state $\rho_0$ such that $E_{H_{\hsh  A}}(\rho_0)<+\infty$.\footnote{The symbol "$\djn$" is defined in (\ref{dj}).}
In particular, if
$$
\lim_{n\to+\infty}E_{H_{\hsh  A}}(\rho_n)=E_{H_{\hsh  A}}(\rho_0)<+\infty
$$
then}
$$
\lim_{n\to+\infty}E_{H_{\hsh  B}}(\Phi_n(\rho_n))=E_{H_{\hsh  B}}(\Phi_0(\rho_0))<+\infty.
$$
\end{property}

\textbf{Note:} The l.h.s. of (\ref{H-dj}) is well defined, since Proposition \ref{dc-l}A in Section 4.2.1
implies that $E_{H_{\hsh  B}}(\Phi_0(\rho_0))<+\infty$.   \smallskip

\emph{Proof.} Since $\Phi_n(\gamma_{H_{\hsh  A\hsh},\beta})=\gamma_{H_{\hsh  B\hsh},\beta'_n}$ tends to $\Phi_0(\gamma_{H_{\hsh  A\hsh},\beta})=\gamma_{H_{\hsh  B\hsh},\beta'_0}$ as $n\to+\infty$,
it is easy to see that
\begin{equation}\label{b-l-r}
 \lim_{n\to+\infty}\beta'_n=\beta'_0.
\end{equation}

We may assume that the r.h.s. of (\ref{H-dj}) is finite and that $E_{H_{\hsh  A\hsh}}(\rho_n)<+\infty$
for all $n\geq0$. It follows that  $\sup_{n}E_{H_{\hsh  A\hsh}}(\rho_n)<+\infty$. So, condition (\ref{Phi-cond}) and Proposition \ref{dc-l}B in Section 4.2.1 imply that
$\sup_{n}E_{H_{\hsh  B\hsh}}(\Phi_n(\rho_n))<+\infty$ and hence
\begin{equation*}%\label{S-conv}
\lim_{n\to+\infty}S(\Phi_n(\rho_n))=S(\Phi_0(\rho_0))<+\infty
\end{equation*}
by the convergence condition for the von Neumann entropy from \cite{W}.

Since
\begin{equation}\label{s-rel}
D(\Phi_n(\rho_n)\|\shs\gamma_{H_{\hsh  B\hsh},\beta'_n})=\beta'_n E_{H_{\hsh  B}}(\Phi_n(\rho_n))-S(\Phi_n(\rho_n))+\ln\Tr e^{-\beta'_nH_B},
\end{equation}
the last limit relation and (\ref{b-l-r}) imply that
\begin{equation}\label{dj-r-1}
\djn(\{D(\Phi_n(\rho_n)\|\shs\gamma_{H_{\hsh  B\hsh},\beta'_n})\})=\beta'_0\djn(\{E_{H_{\hsh  B\hsh}}(\Phi_n(\rho_n))\}).
\end{equation}

Since
$$
D(\rho_n\|\shs\gamma_{H_{\hsh  A\hsh},\beta})=\beta E_{H_{\hsh  A}}(\rho_n)-S(\rho_n)+\ln\Tr e^{-\beta H_A},
$$
we have
\begin{equation}\label{dj-r-2}
\djn(\{D(\rho_n\|\shs\gamma_{H_{\hsh  A\hsh},\beta})\})\leq\beta\djn(\{E_{H_{\hsh  A\hsh}}(\rho_n)\})
\end{equation}
by the lower semicontinuity of the von Neumann entropy.

Proposition \ref{NMP}B in Section 3.2,  (\ref{dj-r-1}) and (\ref{dj-r-2}) imply (\ref{H-dj}). $\Box$ \smallskip

\begin{corollary}\label{H-case-c} \emph{Let $H_X$ be a positive operator on $\H_X$ satisfying condition
(\ref{G-cond}), $X=A,B$. }\smallskip

A) \emph{If $\,\Phi$ is a quantum channel from $A$ to $B$
such that $\Phi(\gamma_{H_{\hsh  A\hsh},\beta})=\gamma_{H_{\hsh  B\hsh},\beta'}$ for some $\beta,\beta'>0$ then
the function $E_{H_{\hsh  B}}(\Phi(\rho))$ is continuous on any subset of $\,\T_+(\H_A)$
on which the function $E_{H_{\hsh  A}}(\rho)$ is continuous.}\smallskip

B) \emph{If $\rho_*$ is a state in $\S(\H_A)$ such that $E_{H_{\hsh  A}}(\rho_*)<+\infty$ then
\begin{equation*}%\label{B-l-r}
\lim_{n\to+\infty}E_{H_{\hsh  B}}(\Phi_n(\rho_*))=E_{H_{\hsh  B}}(\Phi_0(\rho_*))<+\infty
\end{equation*}
for any  sequence  $\{\Phi_n\}$ of channels from $A$ to $B$ strongly converging to a channel  $\Phi_0$
such that $\,\Phi_n(\gamma_{H_{\hsh  A\hsh},\beta})=\gamma_{H_{\hsh  B\hsh},\beta'_n}\,$ for some positive $\beta$, $\beta'_n$, $n\geq0$.}
\end{corollary}\smallskip

\begin{example}\label{Gauss} Let $A$ be one mode quantum oscillator and $\Phi_{N_c, k}$ be the attenuation/amplification/classical noise Gaussian channel
from $A$ to $B=A$  defined by the formula
$$
\Tr\Phi_{N_c, k}(\rho)W(z)=\Tr\rho W(kz)\exp\left[-\frac{1}{2}\left(\!N_c+\frac{|k^2-1|}{2}\right)(x^2+y^2)\right],\quad \rho\in\S(\H_A),
$$
where $z=(x,y)$, $\{W(z)\}$ is the family of Weyl operators (irreducibly acting on the space $\H_A$), $N_c\geq0$ is the power of the environment noise  and $k>0$ is the
attenuation/amplification coefficient \cite[Ch.12]{H-SCI}.

If $\,H_A=\hat{N}\doteq a^*a=\sum_{n\geq0}n|n\rangle\langle n|\,$ is the number operator 
then $\gamma_{\hat{N}\!,\beta}$ is the Gaussian state
$$
\rho(N)=\frac{1}{N+1}\sum_{n\geq0}\left[\frac{N}{N+1}\right]^n|n\rangle\langle n|
$$
with the mean photon number $N=(e^{\beta}-1)^{-1}$. It is easy to show  that
$$
\Phi_{N_c, k}(\rho(N))=\rho(N')
$$
for any $N\geq0$, where $N'=k^2N+\max\{k^2-1,0\}+N_c$ \cite[Section 12.6.3]{H-SCI}.
Thus, the condition of Corollary \ref{H-case-c}A is valid for the channel $\Phi_{N_c, k}$ and this corollary
shows that the channel $\Phi_{N_c, k}$ preserves the local continuity of the function $E_{\hat{N}}(\rho)=\Tr\hat{N}\rho$.

Using Proposition \ref{H-case} with $\Phi_n=\Phi_{N_c, k}$ for all $n$ one can obtain a relation
between discontinuity jumps of the functions $E_{\hat{N}}(\Phi_{N_c, k}(\rho))$ and $E_{\hat{N}}(\rho)$
corresponding to any given converging sequence of states.
\end{example}\smallskip

\begin{example}\label{e-0}
Let $\rho_*$ be a state in $\S(\H_A)$ such that $E_{H_{\hsh A}}(\rho_*)<+\infty$. Let $\{\Phi_t\}_{t\in \mathbb{R}_+}$
be an arbitrary strongly continuous family of quantum channels from $A$ to $B$ such that $\,\Phi_t(\gamma_{H_{\hsh A\hsh},\beta})=\gamma_{H_{\hsh B\hsh},\beta_t'}\,$ for all $\,t\geq0$ and some $\beta,\beta_t'>0$. Then Corollary \ref{H-case-c}B implies that
the functions
$$
t\mapsto E_{H_{\hsh B}}(\Phi_t(\rho_*)),\quad t\mapsto D(\Phi_t(\rho_*)\|\shs\gamma_{H_{\hsh B\hsh},\beta_t'})\quad \textrm{and} \quad t\mapsto S(\Phi_t(\rho_*))
$$
are  continuous on $\mathbb{R}_+$ (the continuity of the last two functions follows from continuity of the first one due to the lower semicontinuity of the QRE and  the relation similar to (\ref{s-rel})).

This observation is applicable to any quantum dynamical semigroup $\{\Phi_t\}_{t\in \mathbb{R}_+}$ preserving the Gibbs state $\gamma_{H_{\hsh A\hsh},\beta}$ (in this case $A=B$ and $\beta_t'=\beta$.)
\end{example}

\pagebreak

\section{Concluding remarks and open questions}
In this article we proved the lower semicontinuity of the function
$$
(\rho,\sigma)\mapsto D(\rho\shs\|\shs\sigma)-D(\Phi(\rho)\shs\|\shs\Phi(\sigma))
$$
for any given quantum operation $\Phi$ and considered its corollaries and applications.
As it turned out, this property allows us to obtain many new local continuity (convergence) conditions
for the QRE and the related functions and to reprove a number of this type results obtained before by different methods.
So, one can say that the established property of the QRE plays a central role in analysis of local continuity of the entropic characteristics
of quantum systems and channels, which either are defined via the QRE (like the QMI) or can be expressed via the QRE (like the mean energy).

Naturally, the question arises about the possibility of proving the same property for others divergence-type functions (the Belavkin-Staszewski relative entropy, the Renyi relative entropy, etc.). Below we describe the basic properties of (the Lindblad extension of the Umegaki) QRE used essentially in the proof of Theorem 1 in Section 3.1:
\begin{itemize}
  \item the lower semicontinuity of the QRE;
  \item the monotonicity of the QRE w.r.t. quantum operations;
  \item the identity $\,D(\rho+\sigma\shs\|\shs \omega+\vartheta)=D(\rho\shs\|\shs \omega)+D(\sigma\shs\|\shs \vartheta)\,$ valid for any
   operators $\rho,\sigma,\omega$ and $\vartheta$ in $\T_+(\H)$ such that $\rho\sigma=\rho\vartheta=\sigma\omega=\omega\vartheta=0$;
  \item the validity of the limit relation
$$
\lim_{n\to+\infty}\textstyle D(P^n_m\rho_nP^n_m\shs\|P^n_m\sigma_n)=D(P^0_m\rho_0P^0_m\shs\|P^0_m\sigma_0)<+\infty\quad \forall m\geq m_0
$$
for any sequences $\,\{\rho_n\}$ and $\{\sigma_n\}$ of operators in $\,\T_+(\H)$ converging, respectively,
to operators  $\rho_0$ and $\sigma_0\neq0$, where $\{P^n_m\}_{n\geq0,m\geq m_0}$ is any double sequence  of finite rank  projectors  completely consistent
with the sequence $\{\sigma_n\}$ (Definition 1 in Section 2.1, Lemma \ref{b-lemma} in Section 3.1);

\item Donald's identity (\ref{Donald});
\item the dominated convergence condition for QRE w.r.t. the first argument: if
$$
\lim_{n\to+\infty}\textstyle D(\rho^1_n\shs\|\sigma_n)=D(\rho^1_0\shs\|\sigma_0)<+\infty
$$
for some sequences $\,\{\rho_n\}$ and $\{\sigma_n\}$  in $\,\T_+(\H)$ converging, respectively,
to operators  $\rho_0$ and $\sigma_0$ then
$$
\lim_{n\to+\infty}\textstyle D(\rho^2_n\shs\|\sigma_n)=D(\rho^2_0\shs\|\sigma_0)<+\infty
$$
for any sequence $\,\{\rho^2_n\}$  in $\,\T_+(\H)$ converging
to an operator  $\rho^2_0$ such that $\rho^2_n\leq c\rho^1_n$ for all $n$ and some $c>0$  (the part of Proposition 2 in \cite{DTL}).
\end{itemize}

The first three of the above properties of the QRE are typical for a divergence-type function, but the last
three of them are quite specific. For example, it seems that the fourth property does not hold for the
Belavkin-Staszewski relative entropy due to its discontinuity in the finite-dimensional settings mentioned in \cite[Proposition 6.7]{C&Co}.

Of course, we do not assert that all the above properties are necessary for the proof
of the analog of Theorem 1 in Section 3.1 for a divergence-type function $D$.

\bigskip

I am grateful to A.S.Holevo, A.V.Bulinski and E.R.Loubenets for useful discussion and comments.
Special thanks to M.M.Wilde for the valuable communication.

\end{document}